%% file: paper.tex
\documentclass[10pt,journal,compsoc]{IEEEtran}
\ifCLASSOPTIONcompsoc
  \usepackage[nocompress]{cite}
\else
  \usepackage{cite}
\fi
\usepackage{color}
\usepackage{times}
\usepackage{epsfig}
\usepackage{amsmath}
\usepackage{amssymb}
\usepackage{caption}
\usepackage{multirow}
\usepackage{graphicx}

\input{comments_tool/macros.tex}

\ifCLASSINFOpdf
\else
\fi
\newcommand{\sonet}{\emph{S2ONet}}
\newcommand{\ovnet}{\emph{O2VNet}}
\newcommand{\vvnet}{\emph{V2VNet}}
\newcommand{\sketchhair}{\emph{DeepSketchHair}}

\begin{document}

%
\title{\sketchhair: Deep Sketch-based \\ 3D Hair Modeling}
%
%
%
%

\author{Yuefan~Shen$^{1}$ \quad Changgeng Zhang$^{1}$ \quad Hongbo Fu \quad Kun Zhou \quad
        Youyi Zheng
        \thanks{$^{1}$ co-first authors}
\IEEEcompsocitemizethanks{
\IEEEcompsocthanksitem Y. Shen, C. Zhang, K. Zhou, and Y. Zheng are with the State Key Lab of CAD\&CG, Zhejiang University, China.\protect\\
E-mail: ~zyy@cad.zju.edu.cn
\IEEEcompsocthanksitem H. Fu is with the School of Creative Media, City University of Hong Kong.\protect\\

}
}

%
%

\markboth{IEEE Transactions on Visualization and Computer Graphics,~Vol.~XX, No.~XX. }
{Shell \MakeLowercase{\textit{et al.}}: Bare Demo of IEEEtran.cls for Computer Society Journals}
%



\IEEEtitleabstractindextext{%
\begin{abstract}
We present \sketchhair, a {deep learning based tool for interactive modeling of 3D hair} from 2D sketches. Given a 3D bust model as reference, our sketching system takes as input a user-drawn sketch (consisting of hair contour and a few strokes indicating the hair growing direction within a hair region), and {automatically} generates a 3D hair model, {which matches the input sketch both globally and locally.}
The key enablers of our system are two carefully designed neural networks, namely,
\sonet, which converts an input sketch to a dense 2D hair orientation field; and
\ovnet, which maps the 2D orientation field to a 3D vector field.
{Our system also supports hair editing with additional sketches in new views. This is enabled by another deep neural network,
\vvnet, which updates the 3D vector field with respect to the new sketches.
}{All the three} networks are trained with synthetic data generated from {a 3D hairstyle database}. We demonstrate the effectiveness {and expressiveness} of our tool using a variety of hairstyles and also compare our method with prior art.
\end{abstract}

}

\maketitle

\IEEEdisplaynontitleabstractindextext

%
\IEEEpeerreviewmaketitle

\section{Introduction}\label{sec:introduction}

The development of digital entertainment leads to the growing demand for 3D content. 3D virtual human plays a critical role in the 3D digital world, while 3D hair is probably the most challenging part of a human body to model. Different from other parts of the human body that can be well modeled as surfaces, the hair is usually modeled as strands, reflecting extreme variability and geometry complexity. What's more, mutual occlusion among strands increases the difficulty of hair modeling.

Image-based hair modeling has gained substantial attention in recent years. It is now possible to {reconstruct} 
realistic and high-quality hairstyles from either multi-view images \cite{zhang2017data,hu2017avatar},  single-view images \cite{chai2012single,chai2013dynamic,hu2015single}, or RGB-D images \cite{zhang2018modeling}. Lately, deep learning based approaches \cite{zhou2018hairnet,saito20183d,zhang2018hair} have also been exploited for modeling 3D hair from images. Although promising results have been achieved with these methods, one major downside in them is that hairstyles are extracted from images rather than free-form designed. For example, all existing image-based techniques focus on extracting unabridged 3D hair from portrait images, and there is no interface {provided} for users to freely modify reconstructed hair models. For consumer-level applications, it is highly desirable to have a system that can not only model hair from images {or even from scratch} but also be capable of performing efficient free-form edits, {to create user-desired hairstyles}.

There are a few prior works focusing on {interactive  modeling of 3D hairstyles from sketches}. Earlier attempts \cite{watanabe1992trigonal,kim2002interactive,choe2005statistical} produce hair models made up of lots of wisps by heavy and complex interaction. The methods of \cite{mao2005sketchy,malik2005sketching,wither2007realistic} aim to simplify interactive modeling of strand-level hair, {which can be achieved by 2D sketches}. However, their generated hairstyles do not own high quality or complex geometry. \cite{yu2001modeling,fu2007sketching} explore vector field based methods and use 3D strokes to drive a 3D vector field which represents the growing direction of hair strands, but {authoring 3D curves is not an easy task for novice users and} it is not intuitive for users to imagine a resulting hairstyle from a vector field. There also exist professional interactive hair modeling systems such as Maya, SolidWorks, 3D Max, CATIA, which incorporate powerful tools for accurate and detailed geometric hair model construction and manipulation. However, to guarantee the quality of produced models, these systems sacrifice usability and often require highly trained skills. 

\begin{figure}[t!]
    \centering
    \includegraphics[width=\linewidth]{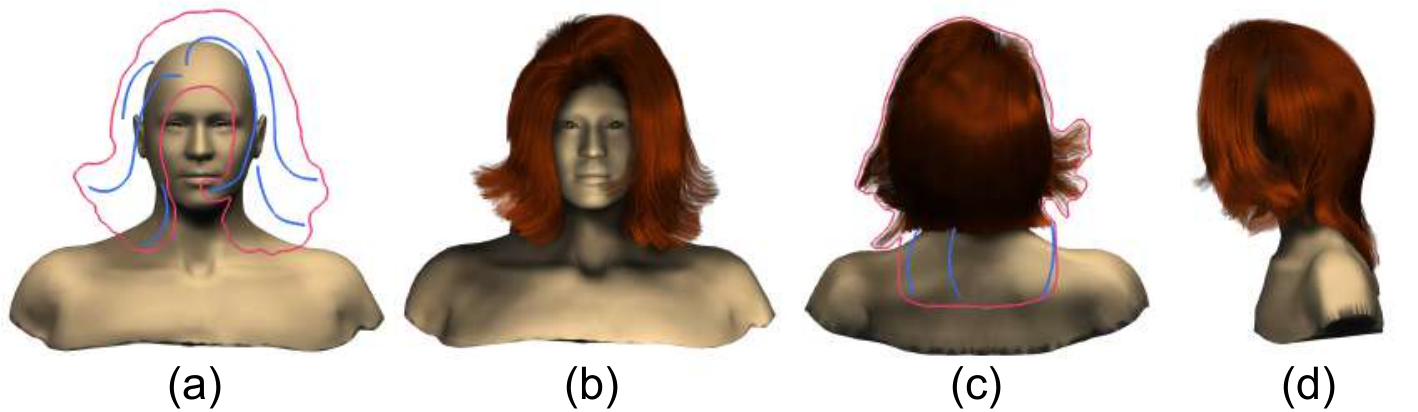}
    \caption{Our system, \sketchhair, takes as input a 2D sketch (a) consisting of a hair contour ({red}) and a few strokes ({blue}) indicating the hair growing directions, and generates a realistic 3D hairstyle (b). Users can continue to modify the generated hair model from another view with a new sketch (c), and our system updates the result accordingly (d).}
    \label{fig:introduction}
\end{figure}

\begin{figure*}
    \centering
    \includegraphics[width=\linewidth]{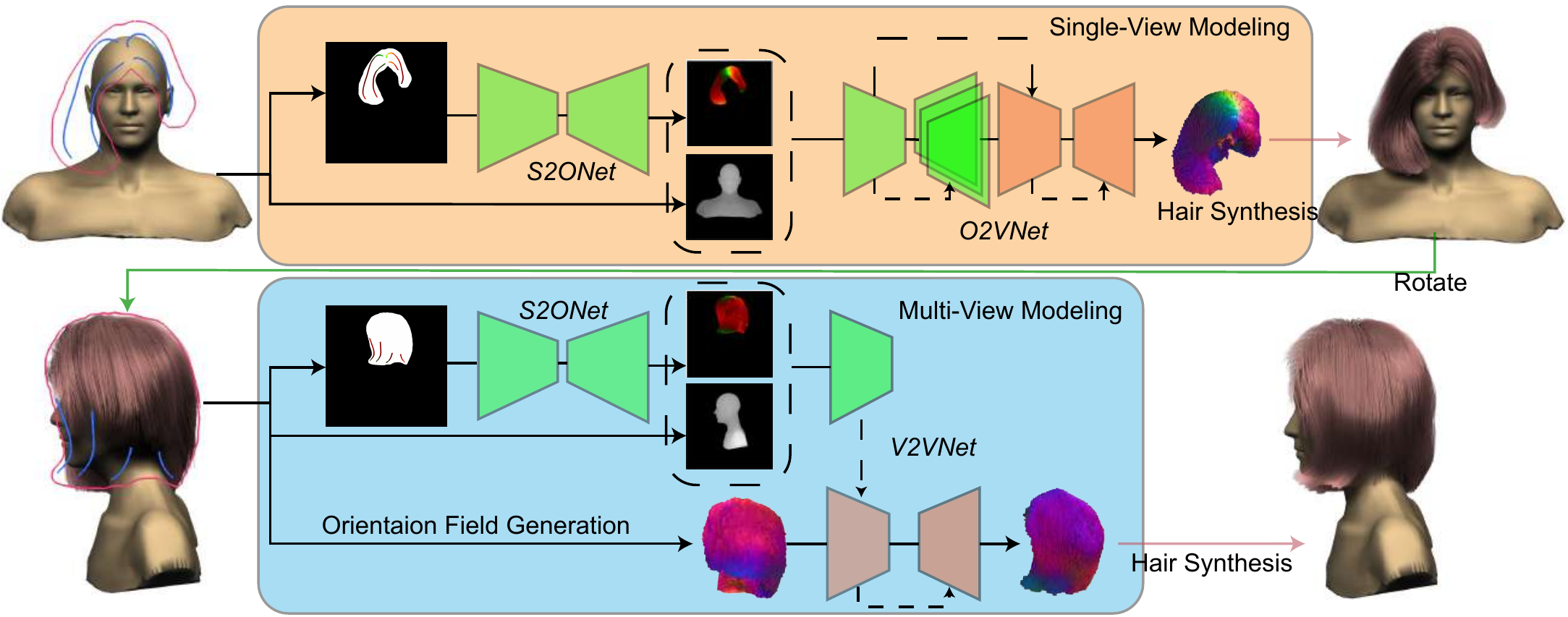}
    \caption{The pipeline of our \sketchhair~framework. The single-view modeling module (top tow) takes as input a user-drawn sketch on top of a bust model (top left), and generates a hair strand model (top right) from {a synthesized intermediate 3D orientation field}. The multi-view modeling module (bottom row) takes as input the bust model, {the currently synthesized hairstyle, and an additional sketch} in a rotated view, and generates the updated strand model (bottom right). Our three generative adversarial networks (\sonet, \ovnet, and \vvnet) are shown in the middle.}
    \label{fig:framework}
\end{figure*}

We introduce \sketchhair, a deep learning based system {for easy, interactive modeling of} high-quality strand-level hairstyles from 2D sketches.
As illustrated in {Fig.
\ref{fig:framework}}, 
our pipeline starts with a {sketched hair contour, which is easily converted to  a {mask map} corresponding to the hair region of a desired 3D hairstyle, and additional strokes as a sketch strand map for rough specification of }sparsely growing direction of the hair. Then \sonet, a generative adversarial network (GAN) model for pixel-to-pixel translation \cite{goodfellow2014generative}, is adopted to predict a dense 2D orientation field, in which each pixel stores the direction of a 2D hair strand passing through it. Next, \ovnet, another GAN-based network, converts the 2D orientation field to a 3D {vector field}\footnote{{We refer to a 3D orientation field as a 3D vector field here to emphasize its difference from a 2D orientation field.}}. To account for the lack of depth information, we add {the depth map of a 3D bust model} as a 3D guidance for \ovnet. Then,  high-quality strand-level hairs are grown from the 3D orientation field. \sketchhair~allows users to modify the resulting 3D hairstyle by inputting additional 2D sketches in new views. Our generated hair models can be free-form adjusted with the help of \vvnet, which is also based on GAN but this time for voxel-to-voxel conversion to update the 3D {vector} field w.r.t. the user input under a new view. Lastly, we provide user-friendly auxiliary tools for fine-tuning the generated hairstyles.

We demonstrate the effectiveness and expressiveness of our modeling system by creating various hairstyles, spanning a diverse spectrum of hair shapes, such as smooth, jagged, wavy, or curly. Our system owns a powerful ability to generate these complex hairstyles, as seen in Fig. \ref{fig:introduction} and Fig. \ref{fig:qualitativ_results}. In summary, our main contributions are listed as follows:
\begin{itemize}
    \item A deep learning based framework for sketch-based hair modeling, which can generate high-quality strand-level 3D hairstyles from 2D sketches;
    \item Three novel generative adversarial neural networks, namely, \sonet,
    \ovnet, and \vvnet, to collaboratively support {interactive multi-view hair modeling.}
\end{itemize}

\section{Related work}\label{sec:Related_work_1}

Our work tackles the problem of interactive hair modeling using deep learning. Below we review existing works closely related to our method, namely, those in hair modeling and sketch-based modeling.

\textbf{Hair Modeling}. Modeling 3D hair strands is a difficult task due to the variety of hairstyles and the complex structures of internal hair. Early works in hair modeling start from scratch while
{recent works focus more on image-based
hair reconstruction.}
We refer readers to~\cite{ward2007survey} and~\cite{bao2018survey} for comprehensive surveys on hair modeling.

Watanabe et al.~\cite{watanabe1992trigonal} introduce a wisp model for 3D hair and this model has been widely applied in early works for interactive hair modeling. Some of these methods directly model hair strands cluster by cluster with drawn 3D curves e.g.~\cite{chen1999system,yang2000cluster,xu2001v}. Other approaches allow users to model a hairstyle roughly first and then solve  wisp-based deformation by interaction~\cite{kim2002interactive,choe2005statistical}. Normally, it is a time-consuming task to model a realistic hairstyle with these methods, and trained skills are required.

{To make the interaction easier}, in~\cite{malik2005sketching} Malik presents a sketching interface for modeling and editing hairstyles by controlling the generation and deformation of hair clusters. Wither et al.~\cite{wither2007realistic} explore a physically-based approach to infer physically-based parameters from 2D user sketches for hair modeling. {Using sketching interfaces makes the modeling process} easier and faster, but the quality of generated hairstyles is not very high.

A vector field based solution~\cite{yu2001modeling} can reduce the time overhead of many manual operations {and supports complex hair modeling} by initializing or editing the {underlying} vector fields. One limitation of this method, however, is that vector fields are not very intuitive for interactive editing. Fu et al.~\cite{fu2007sketching} present a method that allows users to edit a vector field by setting various constraints using 3D curves. However, {authoring 3D curves is not an easy task for novice users and } there is still a gap between {input curves} and generated results.

To reduce the complicated manipulation and strengthen the bond between the input and generated hair, multiple image-based hair modeling methods have been proposed. Ming et al.~\cite{ming1996generation} were probably the first to use real images from different views to generate 3D hair models. Their method builds a 3D volume for hair generation using extracted information from images, like hair outline and hair flow direction. However, this method is only suitable for modeling simple hairstyles. By accurately detecting hair orientation from images, image-based hair modeling is able to generate complex and realistic 3D hair models. Paris et al.~\cite{paris2004capture} use various viewpoints as well as several oriented filters for hair orientation detection in different local parts. Wei et al.~\cite{wei2005modeling} extend image-based hair modeling to a more flexible level, and their method does not rely on special capturing setup under controlled illumination conditions {\cite{paris2004capture}}. More recently, Zhang et al.~\cite{zhang2017data} propose a method which takes only four-view images to generate a 3D hair model using a predefined hair database. Meanwhile, {a few attempts have been made to achieve}
single-view image-based hair modeling. 
For example,~\cite{chai2012single} and~\cite{chai2015high} rely on the human annotation to generate a hairstyle from a single portrait image. Hu et al.~\cite{hu2015single} propose a method based on database retrieval for realistic 3D hair modeling. {Single-view based methods can only generate 3D hair models matching the input view and lack control over hair geometry from new views.}

Deep learning techniques have been exploited for single image-based hair modeling in recent years. Chai et al.~\cite{chai2016autohair} introduce a deep convolutional network for segmenting hair regions and estimating hair growth orientation, which guides their data-driven hair modeling method. {A similar idea is adopted by} Hu et al.~\cite{hu2017avatar} for hair digitization, which uses neural networks for hair classification and region segmentation. Zhou et al.~\cite{zhou2018hairnet} present an encoder-decoder network architecture to generate strand features from 2D orientation input for hair growing. Saito et al.~\cite{saito20183d} extend this idea by adopting a variational auto-encoder structure~\cite{diederik2014auto} to encode the hair information and image information into one latent space. Lately, {Zhang and Zheng}~\cite{zhang2018hair} propose Hair-GAN, which uses a GAN structure with 3D convolutional layers to generate 3D orientation fields for hair growing. {Different from the above approaches, we consider modeling 3D hair from casual 2D sketches. Unlike images that contain rich information, sketches are inherently sparse and ambiguous, which could pose significant challenges for network learning~\cite{li2018robust}.}

\textbf{Sketch-based Modeling}. Sketching is one of the most intuitive and easiest ways for users to interact with computers. However, it remains difficult for computers to interpret freehand sketches due to their inherent sparsity and ambiguity. This leads to a thread of sketch-based modeling methods.

Early free-form modeling methods create 3D contents from user-drawn 2D sketches using inflation (e.g.~\cite{igarashi1999teddy, Mori2007Plushie}). Although inflation is unambiguous, it can only produce rough 3D {surface} models in simple shapes without self-intersections. Further efforts have been dedicated to edit generated rough models by drawing constrain curves like~\cite{Nealen2007FiberMesh}. With strong priors, geometry fitting based approaches retrieve the most similar 3D geometry structures of 2D sketches part by part~\cite{chen20133,shtof2013geosemantic}. Yet, the complexity of their modeling results is limited by simple primitives.

In recent years, deep learning based techniques have also been utilized to learn a desired mapping from 2D sketches to 3D features. Huang et al.~\cite{huang2016shape} introduce deep convolutional neural network (CNN) for mapping 2D sketches to procedural model parameters. Delanoy et al.~\cite{delanoy20183d} propose an end-to-end CNN, which is {trained to generate 3D models from 2D multi-view sketches}. In~\cite{li2018robust}, Li et al. learn 2D middle maps to guide robust 3D modeling from 2D sketches. {Lun et al. \cite{Lun:2017:SketchModeling} generate depth maps from multi-view sketches and then fuse them for 3D reconstruction. Su et al. \cite{Su:2018:ISN} learn normal maps directly from 2D sketches while Han et al. \cite{Han:2017:DDL}  utilize deep neural networks to predict latent code of faces from 2D sketches for the generation of detailed face models.}

Unlike the aforementioned sketch-based modeling methods, which aim to infer 3D surfaces from 2D sketches, our task requires the generation of 3D hair strands exhibiting varying attributes, and thus cannot be solved directly by applying existing sketch-based modeling methods. {To the best of our knowledge,} {our work is the first deep learning based technique for inferring strand-level 3D hairstyles from 2D sketches.}

\section{Overview}\label{sec:Overview}
Fig.~\ref{fig:framework} shows our deep learning based interactive hair modeling pipeline. {Given a bust model, the user sketches over it to scribble a hair mask (the contour, shown in red) and hair strand orientation strokes (shown in blue). We require the user to draw directed strokes to mimic the strand growing direction such that the resulting sketch map is oriented and unambiguous (the tangent direction serves as the orientation at a specific stroke point).} The backbone of our system consists of three deep neural networks {independently aggregating} information to support interactive hair modeling. The first network is named \sonet~(\S\ref{subsec:s2d_network}), which converts the hair mask and the sketch map to a dense 2D orientation field. The second network is named \ovnet~(\S\ref{subsec:d2v_network}), which maps the dense 2D orientation field to a 3D volumetric vector field, {with the help of the depth map of the bust model.
The resulting 3D vector field is then used to grow hair strands for hair synthesis (\S\ref{subsec:hair-strand-growing}). }

Since using only one view of sketches might not be sufficient to generate user-desired 3D hair strands, our approach allows users to adjust the currently synthesized hair strands by drawing additional sketches or masks in new views (\S\ref{sec:multi_modeling}). The third network, named \vvnet~(\S\ref{subsec:network_architec_loss}), then updates the volumetric 3D vector field with respect to the new inputs. More specifically, {a newly drawn sketch is} firstly converted to a dense 2D orientation field (by \sonet), which is then fed into \vvnet~together with the rotated 3D orientation field produced in the previous view to generate the updated 3D orientation field in the current view. The hair strands are updated according to the new orientation field. Users can repeatedly edit the hair strands by sketching in any views.

As complementary tools, auxiliary hair editing operations are added to support hair refinement~(\S\ref{subsec:hair-strand-editing}). The auxiliary operations include cut, deform, color change, enlength, and so on.

\section{Single-view hair modeling}\label{sec:single_modeling}
Given a sketch image $I_S$ and a mask image $I_M$ (Fig. \ref{fig:framework}), our goal is to generate a corresponding 3D vector field or orientation volume $\mathcal{Y}$ to grow realistic hair strands. One primary challenge here is to generate a {desired dense} 3D orientation field from the 2D sparse and inherently ambiguous input. As shown in \cite{li2018robust}, a direct learning of 3D information from  sparse 2D sketches could easily lead to undesirable results. This motivated us to add an intermediate representation, namely, a dense 2D orientation field, between the sketch domain and the volumetric domain to better bridge the gap.

In this section, we focus on our single-view hair modeling framework, which consists of two main parts: \sonet~and \ovnet, as illustrated in Fig.~\ref{fig:framework} (Top). Taking $I_S$ and $I_M$ as input, we first generate a dense 2D orientation field $\Lambda$ with \sonet. Next, $\Lambda$ and the bust depth image $\mathcal{D}$ are taken as the input to generate $\mathcal{Y}$ with \ovnet. Both networks use a conditional GAN structure \cite{pix2pix2017}.

\subsection{\sonet}\label{subsec:s2d_network}
\begin{figure}
    \centering
    \includegraphics[width=\linewidth]{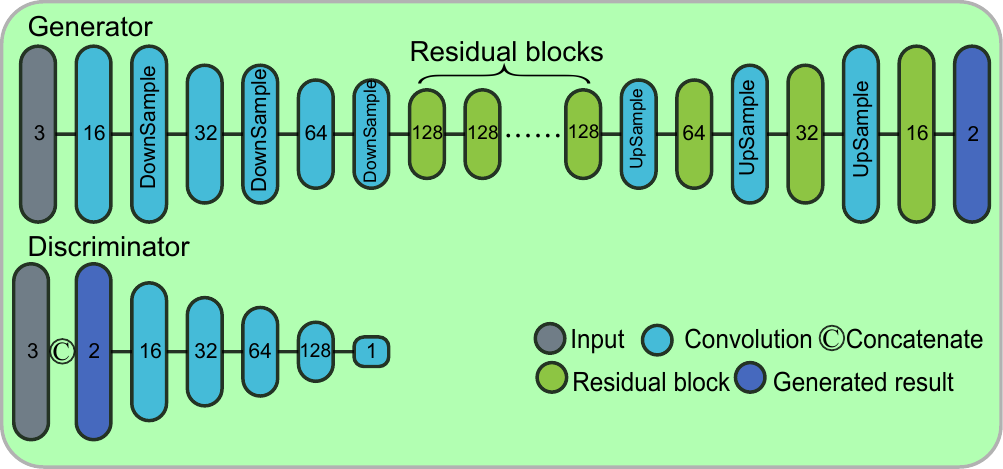}
    \caption{\label{fig:S2DNet} Architecture of our \sonet. For an input pair of sketch image and mask image ${\{I_S, I_M\}}$, we enclasp them into a 2D tensor of size $128 \times 128 \times 3$. For the generator, the contracting part has 3 downsampling convolutional modules with $(32, 64, 128)$ output channels and the expanding part has 3 upsampling deconvolutional modules with $(64, 32, 16)$ output channels connected with 8 residual blocks. The input of our discriminator concatenates the generated \textit{fake} 2D orientation map $\Lambda$ or \textit{real} map $\widetilde{\Lambda}$  with the input condition ${\{I_S, I_M\}}$, which passes through 4 convolutional layers and 1 full-connected linear layer.}
\end{figure}
We apply WGAN-GP~\cite{gulrajani2017improved} for the dense 2D orientation field generation. For this task, the input of the network is a 2D tensor with a size of $128 \times 128$, including 3 feature channels: 2 channels of oriented sketch map and 1 channel of binary mask map. Specifically, the sketch orientation map uses R and G channels in RGB to represent a 2D direction vector $(x,y)$ {(per stroke point and linear mapped from values of [-1,1] to values of [0, 255])}. The output is also a 2D tensor including 2 feature channels, which represents 2D direction vectors $(x,y)$ of the dense 2D orientation field $\Lambda$.

\textbf{Network Architecture.} Fig.~\ref{fig:S2DNet} describes the generator and the discriminator of our GAN network. We use an encoder-decoder structure in the network similar to approaches for generating high-resolution images  with low-resolution inputs~\cite{sangkloy2017scribbler,zhang2018two}. Like~\cite{sangkloy2017scribbler}, our generator consists of residual blocks, upsampling blocks, and downsampling blocks. We also connect the encoder and decoder  with residual blocks.

Since our target dense 2D orientation fields are more unified and simpler than colorful pictures, to avoid overfitting, we simplify the network in~\cite{sangkloy2017scribbler} by reducing half number of filters in all convolutional layers. In addition, our discriminator is composed of five convolutional layers and one linear layer, all without dropout, and takes the corresponding sketch and mask maps as conditional inputs. Specifically, we concatenate the generated fake results or real target with the input sketch and mask images as condition to generate a 2D input tensor with a size of $128 \times 128 \times 5$.

\textbf{Loss Function.} With a sketch image $I_S$ and a mask image $I_M$ as input, we want that our \sonet~generator's result $\Lambda=G_s({I_S, I_M})$ gets close to the target $\widetilde{\Lambda}$.
Previous works on conditional image generation show the benefit of the feature loss, which is defined as the $L2$ difference in a feature space extracted from a middle layer of a pre-trained network \cite{sangkloy2017scribbler}.
However, most of pre-trained networks work on classification or segmentation of photographs while our target images are orientation maps, which are very different from photographs.

Inspired by~\cite{Gatys2016StyleTransfer}, they turn adversarial loss and feature loss to a combination of content loss and style loss. We adopt this idea in our task, and our \sonet~generator loss is defined as:
\begin{equation}
    \begin{aligned}
        \mathcal{L}_{G_s}=\alpha\sum_{l \in m}\mathcal{L}_{content}(l)+\beta\sum_{l \in n}\mathcal{L}_{style}(l)\label{func:L_G_s}.
    \end{aligned}
\end{equation}
Here, the first term is the content loss and the second one is the style loss. $l$ denotes the layer index. {We set $\alpha=0.01$ and $\beta=5$.} We select the middle layers $m=\{0,2\}$ of the discriminator to compute the content loss and middle layers $n=\{0,1,2,3,4\}$ to compute the style loss.
\begin{equation}
    \begin{aligned}
        \mathcal{L}_{content}(l)=\frac{1}{2}\sum_{i,j}[F_{l,i,j}(\Lambda, I_S, I_M)-F_{l,i,j}(\widetilde{\Lambda}, I_S, I_M)]^2.\label{func:L_content}
    \end{aligned}
\end{equation}
In Eq.~(\ref{func:L_content}), $F_{l, i, j}$ represents the discriminator features of the $i^{th}$ filter at position $j$ in the discriminator layer $l$. Note that if $l=0$, we set ${L}_{content}(0)=\frac{1}{2}[\Lambda-\widetilde{\Lambda}]^2$, which means the per-pixel loss between generated results and the ground truth.
Based on~\cite{Gatys2016StyleTransfer}, we define our style loss as follows:
\begin{equation}
    \begin{aligned}
        \mathcal{L}_{style}(l)=\frac{1}{4 N^2_l M^2_l}\sum_{i,j}[&A_{l,i,j}(\Lambda, I_S, I_M)\\
        &-A_{l,i,j}(\widetilde{\Lambda}, I_S, I_M)]^2.\label{func:L_style}
    \end{aligned}
\end{equation}
In Eq.~(\ref{func:L_style}), $A_{l}$ is the Gram matrices, where $A_{l, i, j}$ is the inner product between the vectorized feature maps $i$ and $j$ in the $l^{th}$ layer: $A_{l, i, j}=\sum_k F_{l,i,k}F_{l,j,k}$. $N_l$ is the number of feature channels and $M_l$ is the total size of feature tensors.

The loss of the discriminator is defined as:
\begin{equation}
    \begin{aligned}
    \mathcal{L}_{D_s}= D_s(\widetilde{\Lambda}, I_S, I_M)-D_s(\Lambda, I_S, I_M)+\lambda\mathcal{L}_{gp}, \label{func:L_D_s}
    \end{aligned}
\end{equation}
where
\begin{equation}
    \begin{aligned}
    \mathcal{L}_{gp}=({\Arrowvert \nabla_{\hat{\Lambda}}D_s(\hat{\Lambda}, I_S, I_M)\Arrowvert}_2-1)^2.\label{func:L_gp}
    \end{aligned}
\end{equation}
$D_s(\mathcal{X})$ represents the
\sonet~discriminator. $\mathcal{L}_{gp}$ is the gradient penalty for the random sample $\hat{\Lambda}$ proposed in~\cite{gulrajani2017improved}, where $\hat{\Lambda}=\epsilon \Lambda+(1-\epsilon) \Lambda$, and $\epsilon$ is a random number $\epsilon\sim[0,1]$. {The value of $\lambda$ is set to 10.}

\subsection{\ovnet
}\label{subsec:d2v_network}
\begin{figure}
    \centering
    \includegraphics[width=\linewidth]{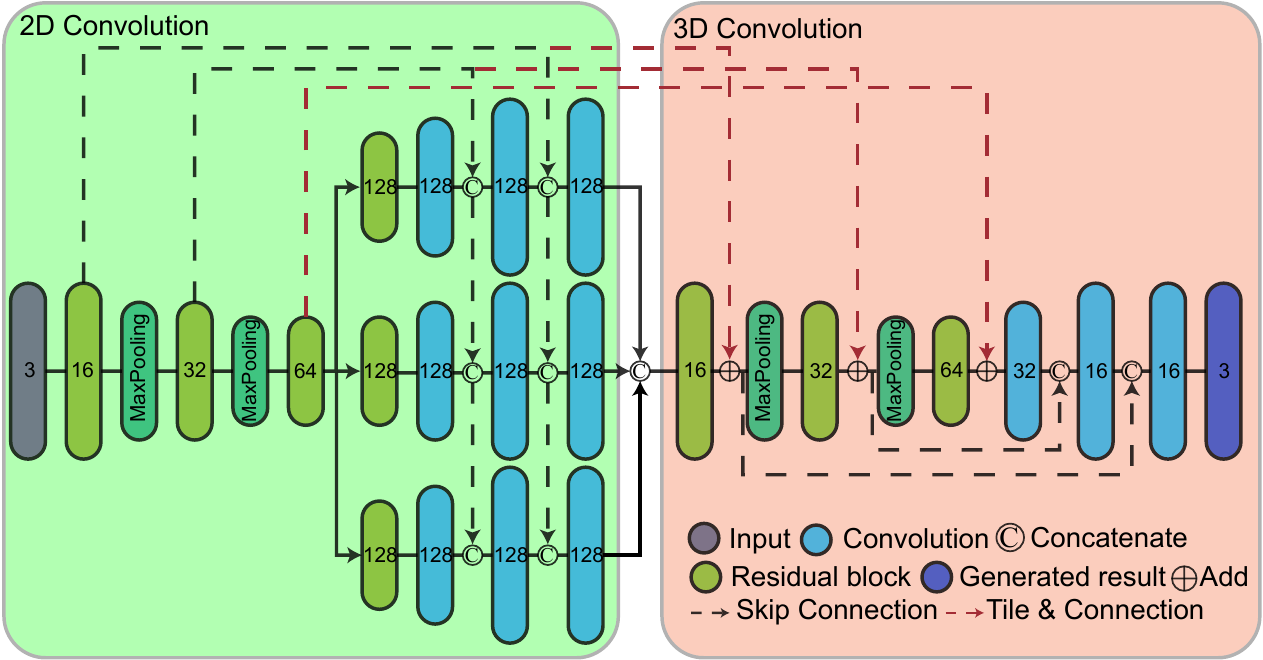}
    \caption{\label{fig:D2VNet}
    Architecture of our \ovnet~generator, which takes the input ${\{\Lambda, \mathcal{D}\}}$ with the shape of $128 \times 128 \times 3$. The encoder of 2D convolutional part takes three residual blocks with $(16, 32, 64)$ output channels combing with two max pooling layers for downsampling, followed by three decoders, each of which contains one residual block and three deconvolutional layers all with $128$ output channels for upsampling. The outputs of three decoders are concatenated and reshaped to a 3D tensor with the size of $128 \times 128 \times 128 \times 3$. The 3D convolutional part resembles a 3D U-Net structure, which includes an encoder with two downsampling modules with $(32, 64)$ output channels and a decoder with two upsampling modules with $(32, 16)$ output channels. The skip connection and our tile \& connection are described in different colors (black and red), respectively.}
\end{figure}
After \sonet, we get a dense 2D orientation field and our \ovnet~is designed to predict a voxel-represented 3D orientation field from this 2D input using a GAN structure. Previous works showed that the GAN with 3D convolutional layers does well in generating voxels \cite{3dgan} and
{has been adopted by Zhang and Zheng~\cite{zhang2018hair}} for image-based hair modeling.
In our case, the input of \ovnet~generator is a 2D tensor with a size of $128 \times 128 \times 3$, including a 2D orientation field $\Lambda$ and a 2.5D bust depth image $\mathcal{D}$. The output is a 3D tensor $\mathcal{Y}$ of size 128 x 128 x 96, representing a 3D orientation field.

\textbf{Network Architecture.} As shown in  Fig.~\ref{fig:D2VNet}, the generator of \ovnet~is constituted by two main {modules}:  a 2D convolutional {module} and a 3D convolutional {module}, each with an encoder-decoder structure based on the commonly used U-Net~\cite{ronneberger2015u}. We share the same discriminator architecture with zhang and zheng~\cite{zhang2018hair}.

The 2D convolutional {module} (Fig.~\ref{fig:D2VNet} (Left)) is composed of an encoder and three decoders of the same shape for predicting values at different axes $X, Y, Z$, similar to~\cite{zhang2018hair}. The output of three decoders can be combined into a 3D tensor of size $128 \times 128 \times 96 \times 3$ from three 2D tensors of {size $128 \times 128 \times 96$}. Unlike~\cite{zhang2018hair}, to make filters in our encoder have larger receptive fields of input images, we set layers of residual blocks all followed by max-pooling layers. Therefore, our \ovnet~has the ability to learn better occupancy features ({see comparisons in $\S$\ref{subsec:experiment_ablation}}).

We notice that using the 2D convolutional module alone cannot generate realistic 3D shapes. Also, if we add simple 3D convolution layers for refinement, the results would lose lots of details originally in 2D orientation maps. To deal with these issues, we build a U-Net using 3D convolution. What's more, we add 3D feature maps with 2D feature maps to ensure that the 2D detail information shall go through the whole network. Note we have to tile 2D feature maps to the size of 3D feature maps before performing feature addition.

\textbf{Loss Function.}
{Like \sonet,}
\ovnet~applies WGAN-GP to speed up the convergence and defines the generator loss as a combination of the content loss and the style loss. In addition, we hope that the projection of the generated 3D orientation field from the input view is as similar as possible to the 2D orientation field to synthesize user-desired hair strands. Thus, we add the projection loss to \ovnet:
\begin{equation}
    \begin{aligned}
    \mathcal{L}_{proj_1}=\sum_{i \in I_{\Theta_1}}(Proj(\mathcal{Y})^{i}-\Lambda_{i})^2\label{func:L_proj_1}.
    \end{aligned}
\end{equation}
Here, $I_{\Theta_1}$ represents the 2D region where {the 2D orientation field has} valid values in the 2D image and $Proj(\mathcal{Y})$ is projected 2D region of the 3D orientation field $\mathcal{Y}$ (from visible voxels). The projection part in our approach differs from existing works, which commonly use a differentiable projection layer for end-to-end training \cite{TulsianiZEM17}. In our case, the camera view of projection is fixed in the training step, and the cross section of the generated voxel shares the same solution as the input 2D orientation field (both are of the size $128 \times 128$). Hence our projection matrix can be fixed. To simplify our training step, we store the indices of the visible voxels in advance, and use the indices to directly compute the value of $Proj(\mathcal{Y})$.

In addition, we add a projection loss to ensure \ovnet~to generate details complying with the input sketch $I_S$.
\begin{equation}
    \begin{aligned}
    \mathcal{L}_{proj_2}=\sum_{i \in I_{\Theta_2}}(Proj(\mathcal{Y})^{i}-I_S^{i})^2,
    \label{func:L_proj_2}
    \end{aligned}
\end{equation}
where $I_{\Theta_2}$ represents the sketch region similar to $I_{\Theta_1}$ and $I_S^i$ is the value at the position $i$ in $I_{\Theta_2}$. 

In our experiments, we find that adding the projection loss could only constrain the voxels in the visible area. To diffuse the constraints to the whole 3D orientation volume, we add the following Laplacian loss:
\begin{equation}
    \begin{aligned}
    \mathcal{L}_{lap}=\sum_{i}(\Delta(\mathcal{Y}_{i})-\Delta(\widetilde{\mathcal{Y}}_{i}))^2,\label{func:L_laplacian}
    \end{aligned}
\end{equation}
with
\begin{equation}
    \begin{aligned}
    \Delta(\mathcal{Y}_{i})=\sum_{j \in N_i}\frac{1}{|N_i|}(v_j - v_i).\label{func:L_lap_ope}
    \end{aligned}
\end{equation}
Our Laplacian loss represents the difference of the divergence of the gradient between the generated 3D orientation field and the ground truth, {denoted as $\widetilde{\mathcal{Y}}$}. $\Delta(\mathcal{X})$ is the Laplace operator and $i$ in Eq.~\ref{func:L_laplacian} represents each voxel in the 3D orientation field. In Eq.~\ref{func:L_lap_ope}, $N_i$ is the index set of the neighbors of the voxel $\mathcal{Y}_i$, and $v_i$, and $v_j$ are the values of the voxels $\mathcal{Y}_i$ and $\mathcal{Y}_j$, respectively.

Our final generator loss is defined as:
\begin{equation}
    \begin{aligned}
    \mathcal{L}_{G_d}=&\iota \mathcal{L}_{content}+\kappa \mathcal{L}_{style}\\
    &+\gamma \mathcal{L}_{proj_1}+\delta \mathcal{L}_{proj_2}+\varepsilon \mathcal{L}_{lap}. \label{func:L_all}
    \end{aligned}
\end{equation}
In our experiments, $\iota = 0.01$, $\kappa = 5$, $\gamma = 0.1$, $\delta = 0.5$, and $\varepsilon = 2e-5$. Note we do not normalize each loss.

\subsection{Data Generation}\label{sec:single_view_data_generation}
{To train our networks, we prepared a hair dataset, which is} partly from public online repositories \cite{saito20183d} and partly provided by \cite{chai2016autohair}. We collected totally 653 3D {strand-level models}, and aligned each hair model to a unified bust model within a volumetric bounding box. We fixed the camera at the front of the box and placed a view-orthogonal sketch plane in the middle of the box.

For our single-view modeling, we project each {pair of hair and bust models} to the front view to generate the rendering of the corresponding ground-truth 2D orientation map, mask map, and bust depth map. We randomly rotate the hair-strand model and bust model together around the box center to augment the training dataset. The rotation ranges from $+30^o$ to $-30^o$ for the Y axis and $+15^o$ to $-15^o$ for both the X and Z axes. {X, Y, and Z are the world coordinate axes}. We describe some details in the followings.

\textbf{2D Hair Maps.}
To generate dense 2D orientation maps, we render hair strands into images with every vertex's color of strand representing its tangent direction. {We use OpenGL depth test to avoid rendering of invisible strands}. Similarly, we generate hair mask maps by replacing every strand vertex's color {in rendered orientation maps} by white. To generate the depth map of the bust model, we cast a ray parallel to the Z axis, assuming the ray intersects the bust in $p$ and the z-component value of $p$ is $p^{z}$. We define $b_{min}$ as the box minimum point whose depth value $b_{min}^{z}$ is the minimum among all box points. The box maximum point $b_{max}$ and its depth value $b_{max}^{z}$ are similarly defined. Then the {normalized} depth of the current ray intersection is calculated as $\frac{p^{z} - b_{min}^{z}}{b_{max}^{z} - b_{min}^{z}}$. Finally, we get the dense orientation map $\Lambda$, the mask map $I_M$, and the bust depth map $\mathcal{D}$. All of the hair maps have the resolution of $128 \times 128$ in our implementation. {Unlike \cite{zhang2018hair}, which uses a high resolution image ($1024\times1024$) and then performs downsampling to $128 \times 128$, we use a direct $128 \times 128$ input for two reasons. On one hand, such a resolution is in {conjunction} with the fused 3D volume and in our experiments we find it to be sufficiently informative to synthesize vivid 3D orientation fields for hair synthesis (see comparisons to \cite{zhang2018hair} in  \S \ref{sec:experiment}); On the other hand, synthesizing high resolution images with high quality is known to be difficult for a GAN model such as our \sonet~\cite{chen2017photographic}.

\begin{figure}
    \centering
    \includegraphics[width=\linewidth]{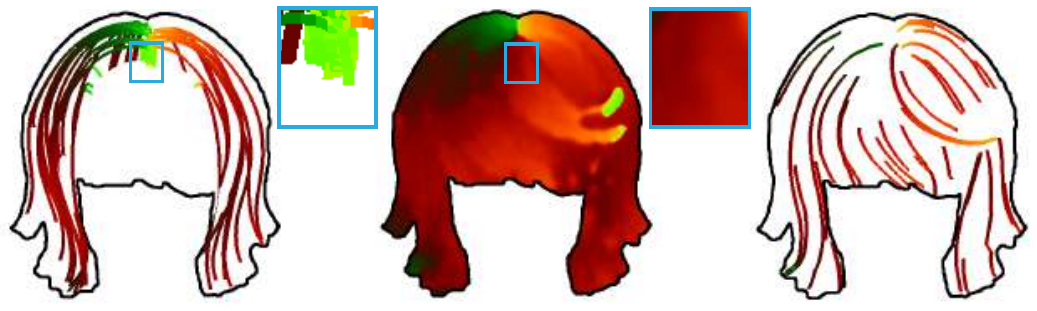}
    \caption{From left to right: the sketch map generated by rendering randomly selected hair strands, the corresponding ground-truth 2D orientation map, and the sketch map generated with our image tracing method. The rendered result (Left) may include orientation information from invisible hair strands, and does not match with the GT 2D orientation map well. In additional, a random selection may lead to non-uniform distribution.}
    \label{fig:sketch_compare}
\end{figure}

\textbf{2D Sketch Map.} A direct way to generate a sparse sketch map for training \sonet~is to randomly sample hair strands from an entire hair model. However, one might note that the dense 2D orientation map we wish to synthesize only depicts the hair orientation of visible strands. Sampling from invisible hair strands could lead to ambiguities in the input, as shown in Fig. \ref{fig:sketch_compare} (Left). To avoid such ambiguities, we generate the sketch map by tracing the dense 2D orientation map. Specifically, we first randomly select {a seed pixel} on the dense orientation map, and {trace from this seed pixel by iteratively finding} the next pixel among eight neighboring pixels with the smallest color discrepancy. Supposing the current pixel value is $p(x,y)$ ({representing the projected strand orientation at this position}) and the value for one of its neighbors is $p^{n}(x,y)$, we mark $p_{n}$ as a candidate pixel if the dot product between $p(x,y)$ and $p^{n}(x,y)$ is below a given threshold ($\varrho = 0.5$ in our case). After checking all eight neighboring pixels of the selected pixel, we choose the most matching pixel among the candidate pixels as the next iterative pixel by finding the minimum dot product. If there is no candidate pixel, we terminate the tracing process and {randomly} re-select a new seed pixel for the next tracing process. Finally, we get the sketch map $I_S$ (Fig. \ref{fig:sketch_compare} (Right)) through multiple iterations. {To ensure uniform sampling, we employ an adaptive clustering strategy as in \cite{Wang2009EHG} and then select one strand in each cluster.}

\textbf{3D Hair Orientation Field.}  As in~\cite{zhang2018hair}, we convert hair strands to a 3D orientation field $\mathcal{Y}$. We define a grid volume with the size of $ 128 \times 128 \times 96 $. Every grid {cell} stores a value which is the average tangent direction of hair strands that pass through this grid {cell}.

\section{Multi-view hair modeling}\label{sec:multi_modeling}
In our system we allow users to draw a new sketch or mask to adjust the hair model {in either the current view or a novel view.} This is fully supported by our multi-view hair modeling module. This module takes as input the 3D orientation field $\mathcal{Y}$ (after rotation), the bust depth map (after rotation), and the newly specified sketch and hair mask. We first generate a dense 2D orientation field {$\Lambda^*$} from the new sketch and mask images using \sonet. The system then outputs a new 3D orientation field $\mathcal{Y}^*$. In this process, we hope that our network only changes the 3D orientation field in the new view {with respect to the new sketch}, while it can keep the original values in the other areas. Existing works on multi-view modeling like~\cite{delanoy20183d} simply concatenate 3D volume with 2D feature maps and feed them to a network only using 2D convolutional layers. However, through preliminary experiments, we find this structure performs badly for our task {(\S\ref{sec:experiment})}. We thus design a volume-to-volume network structure guided by 2D feature maps, named \vvnet.

\subsection{\vvnet~Architecture and Loss Function}\label{subsec:network_architec_loss}
\begin{figure}
    \centering
    \includegraphics[width=\linewidth]{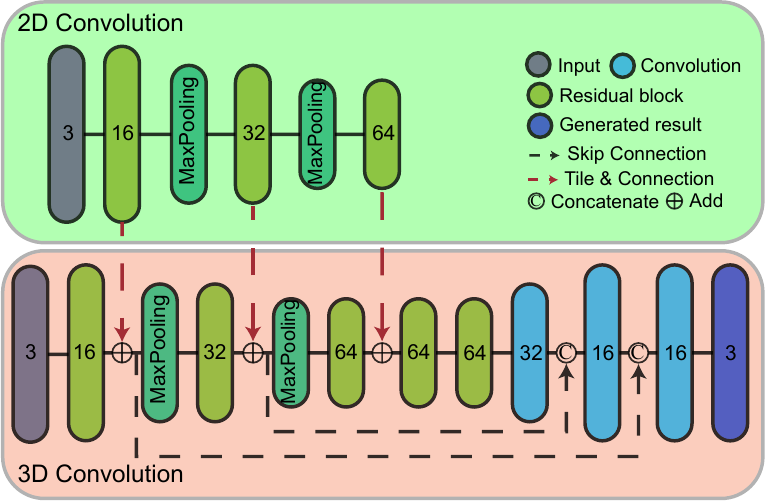}
    \caption{\label{fig:V2VNet}
    The generator architecture of our \vvnet. The structures of 2D and 3D convolution are similar to that of \ovnet.}
\end{figure}
The \vvnet~architecture is similar to the 3D convolutional module in \ovnet, and is based on the V2V-PoseNet~\cite{moon2018v2v}. It takes a rotated 3D orientation field $R(\mathcal{Y})$ where $\mathcal{Y}$ is the orientation field in the previous view, a 2D orientation map $\Lambda^*$ and a bust depth image $\mathcal{D}^*$ as input and generates an updated 3D orientation field $\mathcal{Y}^*$. As shown in Fig.~\ref{fig:V2VNet}, \vvnet~has two modules. The 2D convolutional part takes a new 2D orientation map {$\Lambda^*$} and a new bust depth image $\mathcal{D}^*$ as input and applies 2D convolutional layers with 3 residual blocks and 2 downsampling layers to extract 2D feature maps for guiding the volume-to-volume generation. The 3D convolutional part uses a U-Net structure, with both the encoder and the decoder consisting of residual blocks. Unlike the 3D convolutional sub-network in the \ovnet, we add more residual blocks in our \vvnet~because this task is to modify the input 3D orientation field. All the convolutional layers except the last layer in our three networks are followed by a ReLU activation.

\textbf{Loss Function.} \vvnet~is also based on WGAN-GP, and has the same discriminator as before. Meanwhile, for the generator of \vvnet, we add a loss to help the network keep the original features of the pre-generated 3D orientation fields:
\begin{equation}
    \begin{aligned}
    \mathcal{L}_{ori}=\sum_{i \in \Gamma}(\mathcal{Y}^*_i-R(\mathcal{Y})_i)^2.\label{func:L_ori}
    \end{aligned}
\end{equation}
Here, $\Gamma$ represents the
{invisible hair voxels} in the 3D orientation field from the current view. Similar to $I_{\Theta_1}$ and $I_{\Theta_2}$, we can initialize an index set of such invisible voxels in advance. The full loss of the \vvnet~generator is defined as:
\begin{equation}
    \begin{aligned}
    \mathcal{L}_{G_v}=&\iota \mathcal{L}_{content}+\kappa \mathcal{L}_{style}+\gamma \mathcal{L}_{proj_1}\\
    &+\delta \mathcal{L}_{proj_2}+\varepsilon \mathcal{L}_{lap}+\zeta \mathcal{L}_{ori}.\label{func:L_g_v}
    \end{aligned}
\end{equation}
{Here {the values of weight parameters $\iota, \kappa, \gamma, \delta, \varepsilon$ are same as those used in Eq. \ref{func:L_all}} and we set $\zeta = 0.1$.}

\subsection{Data Preparation}\label{subsec:data_preprocess}
\begin{figure}
    \centering
    \includegraphics[width=\linewidth]{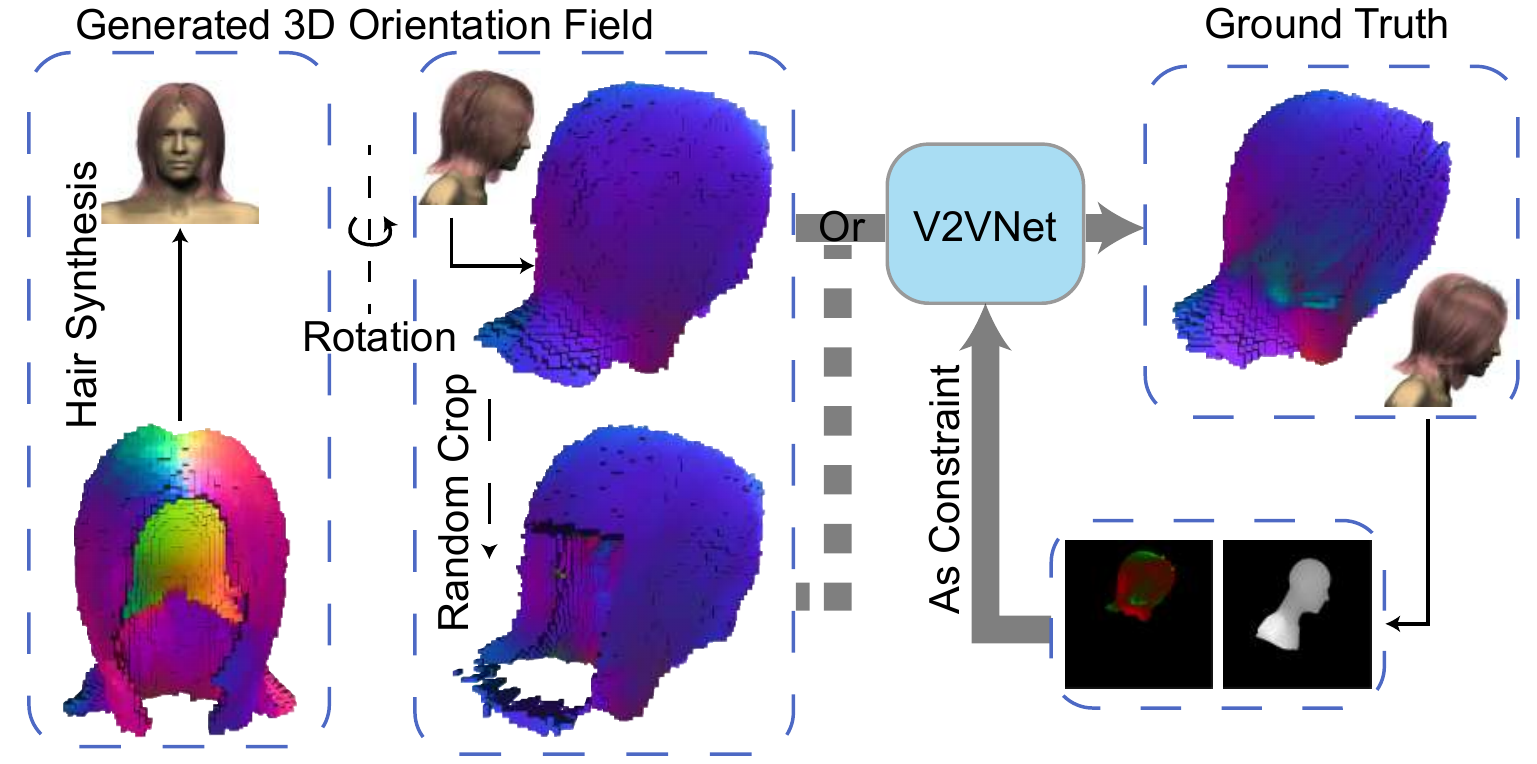}
    \caption{\label{fig:multiviewdatapre}
    Training data generation for \vvnet. Left: we use the generated 3D orientation field $\mathcal{Y}$ from \ovnet~to synthesize hair strands. Middle: The rotated 3D hair model is then used for generating the rotated 3D orientation field $R(\mathcal{Y})$ as the training input for \vvnet. The middle bottom shows randomly cropped $R'(\mathcal{Y})$. Our \vvnet~takes $R(\mathcal{Y})$ or $R'(\mathcal{Y})$ as the input and uses the 2D orientation map and bust depth map generated from the paired ground truth (right) as constraints.}
\end{figure}
As mentioned earlier, the input and the output of \vvnet~are different 3D orientation fields, but our current dataset contains 2D hair maps, bust depth maps, and 3D orientation fields from the front view only. To train \vvnet, we require paired data $((R(\mathcal{Y}),\mathcal{D}^*,\Lambda^*), \mathcal{Y}^*)$. Here, $\mathcal{D}^*$, $\Lambda^*$, and $\mathcal{Y}^*$ are readily available from the current view. However, the input 3D orientation field $R(\mathcal{Y})$ is missing, which indicates the rotated orientation field from the previous view and should merit slight differences to $\mathcal{Y}^*$ since we are expecting $\mathcal{Y}^*$ to be an updated 3D orientation field matching the 2D orientation map $\Lambda^*$.

Profiting from our single-view hair modeling, the synthesized 3D orientation fields naturally satisfy the above requirements. Thus, we use our 2D orientation maps in the front view and pre-train \ovnet~to generate 3D orientation fields $\mathcal{Y}$. And then, we rotate $\mathcal{Y}$ to a random angle imitating user editing. {However, the rotation of discrete grids or volumes over a continuous space will bring in aliasing.} Interpolation is required to reduce the aliasing of the rotation. In our case, we resort to an alternative solution. First, we synthesize the hair strands using the 3D orientation field $\mathcal{Y}$. We then rotate the hair strands and calculate the new 3D orientation field $R(\mathcal{Y})$ {from the rotated hair strands}. We will describe our hair synthesis implementation using 3D orientation fields later in \S\ref{subsec:hair-strand-growing}. In addition, we randomly crop some areas of $R(\mathcal{Y})$ from the current view. This operation makes $R(\mathcal{Y})$ have a different silhouette from the input mask. In total, there are 5,000 and 50 samples for training and validation, respectively. Fig.~\ref{fig:multiviewdatapre} illustrates the process.

\subsection{Hair Synthesis}\label{subsec:hair-strand-growing}
\begin{figure}
	\centering
	\includegraphics[width=\linewidth]{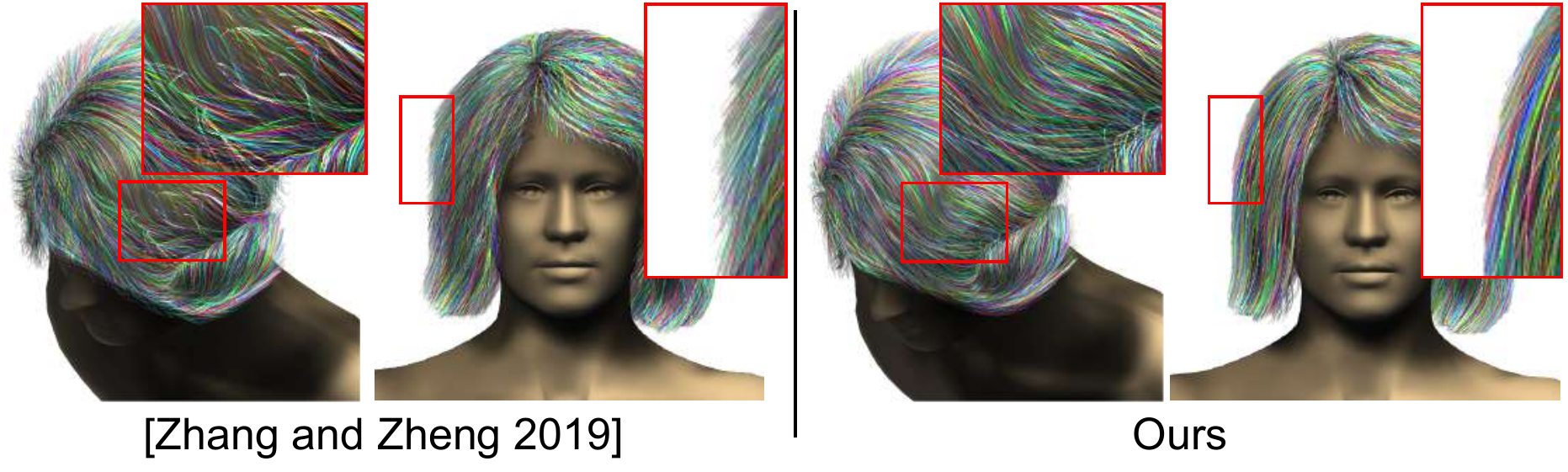}
	\caption{\label{fig:growing_compare}Comparison of our hair growing algorithm to that by Zhang and Zheng~\cite{zhang2018hair} given the same 3D orientation field.}
\end{figure}
In Section  \ref{sec:single_view_data_generation}, we introduce how to convert strand-level models to voxel-level models. We now introduce the method that converts voxel-level models to strand-level models. As mentioned above, we get a 3D orientation field from {the output of either \ovnet~or \vvnet}~with every grid cell storing a value, which is the average tangent direction of hair strands that pass through this cell.


Our hair growing algorithm is similar to that of \cite{chai2013dynamic} and \cite{zhang2018hair}. We do some {modifications} to reduce the influence of noisy grids. First, we start from the scalp and grow hair strands from the root positions as in \cite{chai2013dynamic}. A strand will continue to grow along the grid cell direction to reach its next grid cell unless the next grid cell is out-of-volume or the angle difference between its direction $d$ and the previous direction $\bar{d}$ is larger than a threshold $\theta$. {Unlike \cite{chai2013dynamic} and \cite{zhang2018hair}, which stop the growing with a small value of $\theta$, we set $\theta$ to $150^o$. More importantly, in case when $\theta$ larger than} $60^o$, we set the new growing direction to be the mean direction between $d$ and $\bar{d}$ to ensure smoothness.

The above process gives us a set of ``good" hair strands, denoted as $S_g$. We then randomly select seed grid cells inside the hair volume which have not been passed by $S_g$ and whose value is valid {(a grid value $(x,y,z)$ is considered to be valid if $(x\times{x}+y\times{y}+z\times{z}) \geq 0.5$).} Then we perform the growing algorithm as in \cite{zhang2018hair}. This process might result in candidate hair strands $S_c$ that may not connect to hair roots. To address this issue, for each strand $s^i \in S_c$ we try to find the most similar strand $s^j$ from $S_g$,
with the similarity between $s^i$ and $s^j$ measured in terms of their curvature. If more than $1/3$ of the continuous strand vertices of $s^i$ whose curvature matches with that of $s^j$, we set $s^j$ as the guiding hair strand for $s^i$ and continue growing $s^i$ using the direction of $s^j$ and finally connect it to the neighboring roots of $s^j$ if available. {Fig. \ref{fig:growing_compare} shows a comparison of our hair growing algorithm to that of \cite{zhang2018hair}. Note that our hair strands are more continuous and venerable in noise regions.}

\section{User Interface}\label{sec:hair_synthesis}
\begin{figure}
	\centering
	\includegraphics[width=\linewidth]{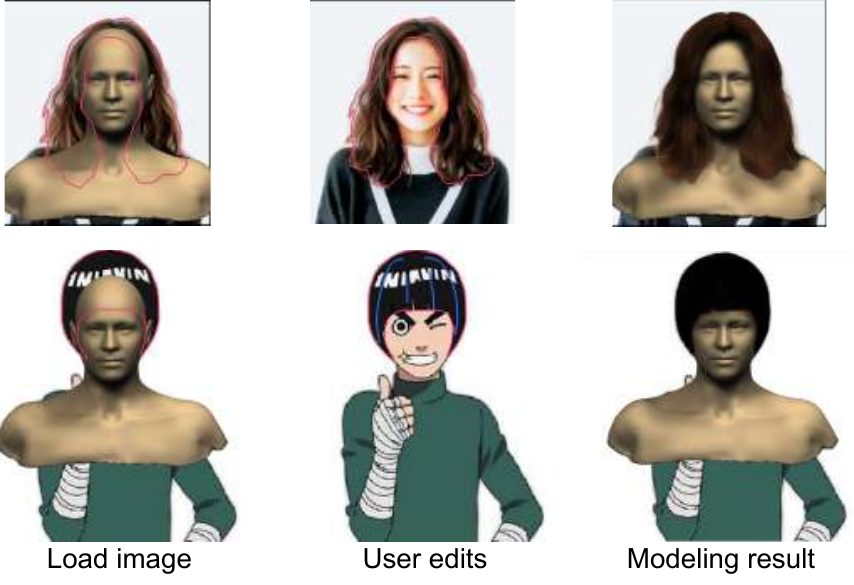}
	\caption{The user interface. The user can load an existing portrait image as a reference or s/he can start from scratch as shown in Fig. \ref{fig:introduction}. The hair masks and orientation maps can be automatically computed from the images by the method of \cite{chai2016autohair}.}
	\label{fig:user_interaction}
\end{figure}


\begin{figure}
	\centering
	\includegraphics[width=\linewidth]{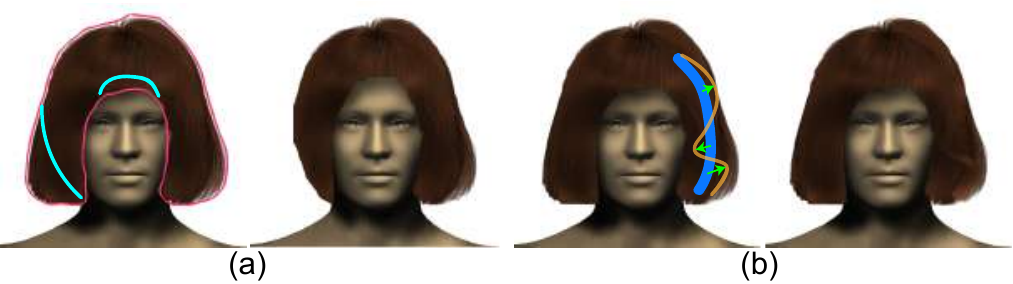}
	\caption{Auxiliary tools ((a) cut and (b) local deform) supported in our system.}
	\label{fig:aux_ops}
\end{figure}

Our system allows a user to {design 3D hairstyles with or without a reference. A hairstyle can be created from scratch. In this case}, the user can directly draw strokes and hair mask atop a 3D bust model (initially in the front view). Then the user-specified sketch map and the mask map are sent to our networks to produce a 3D orientation field for hair synthesis. Alternatively, we allow the user to import a portrait image to depict a desired hairstyle (Fig. \ref{fig:user_interaction}). In this case, our system will automatically align the image to the bust model by using the face alignment algorithm proposed by Cao et al. \cite{cao2014displaced} and segment the hair mask map by the method in \cite{chai2016autohair}. {The orientation map is also automatically extracted from the input image using the method described in \cite{zhang2018hair}. In case when the user draws orientation strokes, we apply our \sonet~to generate the 2D orientation map instead.} The generated maps are then fed into our networks to produce a 3D orientation field. Such referencing images can be used in other views as well. {For multi-view modeling, the user performs the same set of operations to specify the new hair mask and strand orientations.}

\subsection{Hair Strand Editing}\label{subsec:hair-strand-editing}
We propose some strand-level hair model editing tools for users to do simple hairstyle modifications. {The supported tools include hair cutting, reshaping, lengthening(by scaling), color changing, and texturing.} {Below we give the details on strand cutting and reshaping.}


\textbf{Strand Cutting.} As illustrated in Fig. \ref{fig:aux_ops} (a), this tool is used to cut the hair and trim the boundary of a hair strand model. The user can directly cut 3D hair strands by 2D strokes. We identify hair strands which intersect with the user-specified strokes, and discard the strands' portion which is not connected to the root. Alternatively, the user can adjust the hair strand model by editing the 2D hair mask. Our system monitors the 2D contour of the hair strand model and computes the new hair mask, which is then mapped to the 3D hair volume. Then we discard hair strand vertices which are outside the modified hair volume. {Note when the user enlarges the hair volume by re-sketching the hair mask (i.e., not a cutting operation), we need \ovnet~and \vvnet~to update the 3D orientation field.}

\textbf{Strand Reshaping.} As illustrated in Fig. \ref{fig:aux_ops} (b), this tool is used to deform a wisp of hair strands. Our interface allows the user to choose a wisp of hair strands in two ways. First, since a wisp of hair strand must grow from adjacent hair roots, the user can pitch on the scalp area and all hair strands whose roots are in the selected area will be selected. In the other way, the user can select a wisp of hair strands by drawing a strand sketch on the sketch plane and our system then selects hair strands whose projected shape matches the strand sketch by measuring the vertex curvatures.
The selected wisp of hair strands can be deformed simply by moving the selected strand vertices and the rest of the strand vertices are updated using a method based on Laplacian editing \cite{Sorkine:2004:LSE}.

\begin{figure*}
	\centering
	\includegraphics[width=\linewidth]{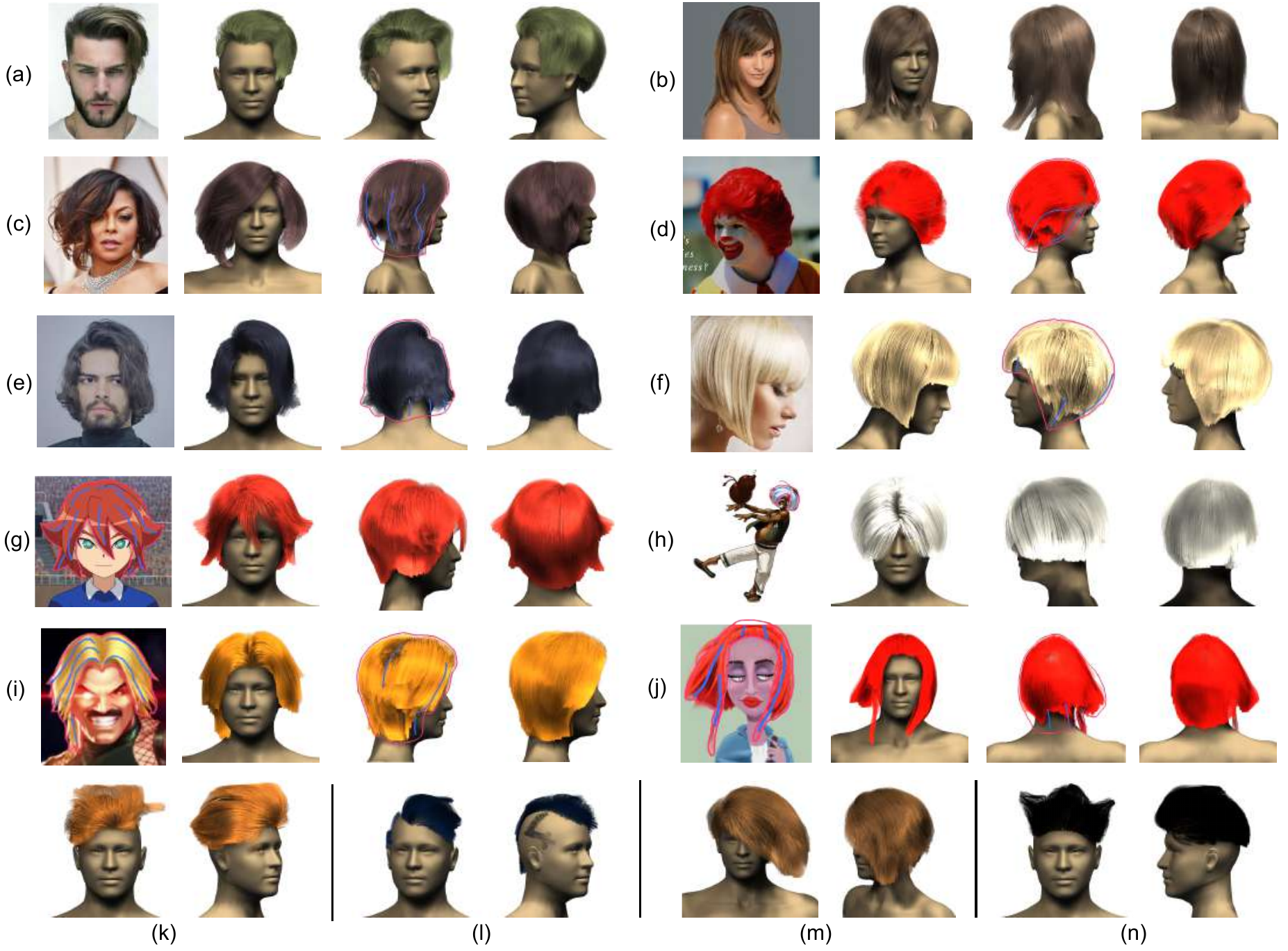}
	\caption{A gallery of modeling results of 3D hairstyles from different inputs. Hairstyles in (a) and (b) are modeled from single images automatically. From left to right in (c) to (j), we show the reference images, the generated results, results in a different view, and the results after further edits {by multi-view modeling (for g and h, no further edits were performed)}. The hairstyles created from scratch are shown in (k), (l), (m) and (n). {The input sketches if any are overlaid with the reference images.} }
	\label{fig:qualitativ_results}
\end{figure*}

\section{Experimental results}\label{sec:experiment}
Our interactive system is implemented using QT 5.12 and OpenGL. Our three networks (\sonet, \ovnet, and \vvnet) are implemented using the TensorFlow framework and trained with 4 GeForce GTX 1080Ti GPUs. It took 100 epochs for \sonet~and 200 epochs for the other networks all with a batch size of 8 for training. Note that, for the first 100 epochs of \ovnet~and \vvnet~training, we do not use the projection loss or Laplacian loss. This training strategy makes the training converge faster. The whole training process took about one week. In the runtime, all of our experimental tests are conducted on a PC with an Intel(R) Core(TM) i7-8770 3.20GHz CPU and 16GB memory. Meanwhile, our system needs only one GPU for the feedforward propagation of the three networks.

\subsection{Performance}\label{subsec:experiment_performance}

Since our target is to model high quality 3D hairstyles with 2D sketches, the performance of our \sketchhair~system is recorded in the following two aspects.

\begin{table}
 \renewcommand{\arraystretch}{1.3}
 \centering
 \begin{tabular}{|c|c|c|c|c|c|c|c|}
  \hline
  \bfseries Index & a & b & c & d & e & f & g\\
  \hline
  \bfseries views & 1 & 1 & 2 & 2 & 2 & 2 & 1\\
  \hline
  \bfseries strokes & 0 & 0 & 0, 3 & 0, 3 & 0, 3 & 0, 3 & 10\\
  \hline
  \hline
  \bfseries Index & h & i & j & k & l & m & n\\
  \hline
  \bfseries views & 1 & 2 & 2 & 3 & 2 & 2 & 2\\
  \hline
  \bfseries strokes & 5 & 6, 3 & 5, 2 & 7, 3, 2 & 3, 0 & 4, 2 & 5, 2\\
  \hline
 \end{tabular}
 \caption{Statistics of interaction for results shown in Fig.~\ref{fig:qualitativ_results}. The stroke numbers are in accordance with the number of views.}
 \label{tab:stat}
\end{table}

\textbf{Qualitative Results.} We have evaluated the performance of our \sketchhair~system on some Internet portrait images and cartoon images, {for which it is challenging for frontier image-based hair modeling methods to deal with}. Note that, our system can generate 3D hair models from portrait images automatically by first using the method in~\cite{chai2016autohair} to generate hair mask images and dense 2D orientation maps. Meanwhile, users are allowed to perform freeform edits atop the creation. The hairstyles in all these cases are of various shapes and details. Some of the results are shown in Fig.~\ref{fig:qualitativ_results}, and~\ref{fig:user_results}. {In Fig.~\ref{fig:qualitativ_results}, some of the examples are not taken in frontal view (e.g., (d), (f), and (h)), some of them are of low resolution (e.g., (h)) while some of them are in cartoon format in which the hairstyles are dramatically different from real hairs ((g) - (j)). All aforementioned issues pose potential challenges for existing single-view image-based techniques as they either require frontal view or high-resolution input images, or rely on existing hair databases \cite{saito20183d,zhang2018hair}.} The 3D hair models in the first row of Fig.~\ref{fig:qualitativ_results} are generated automatically while the 3D hairstyles {in the second to fifth rows are obtained after {inputting additional sketches in new views} to get more vivid results. The last row shows examples of freeform hairstyles that are created from scratch.} Each hairstyle took less than 2 minutes to model by a trained user. The statistics including the number of views and the number of strokes to produce the associated results are shown in Table \ref{tab:stat}. Note our single-view modeling does not require a frontal view to start (e.g., Fig. \ref{fig:qualitativ_results}, f) since we have training data from distant views (\S \ref{sec:single_view_data_generation}).

We have the following findings from the qualitative results: 1) Our \sketchhair~system can generate state-of-the-art image-based hairstyles; 2) The generated results are highly realistic, no matter if the reference images are unreal cartoon pictures or real portrait images. 3) Our system is capable of generating various crazy hairstyles such as those in the last row of Fig.~\ref{fig:qualitativ_results}. All of these demonstrate the {effectiveness and expressiveness} of our system.

\begin{figure}
	\centering
	\includegraphics[width=\linewidth]{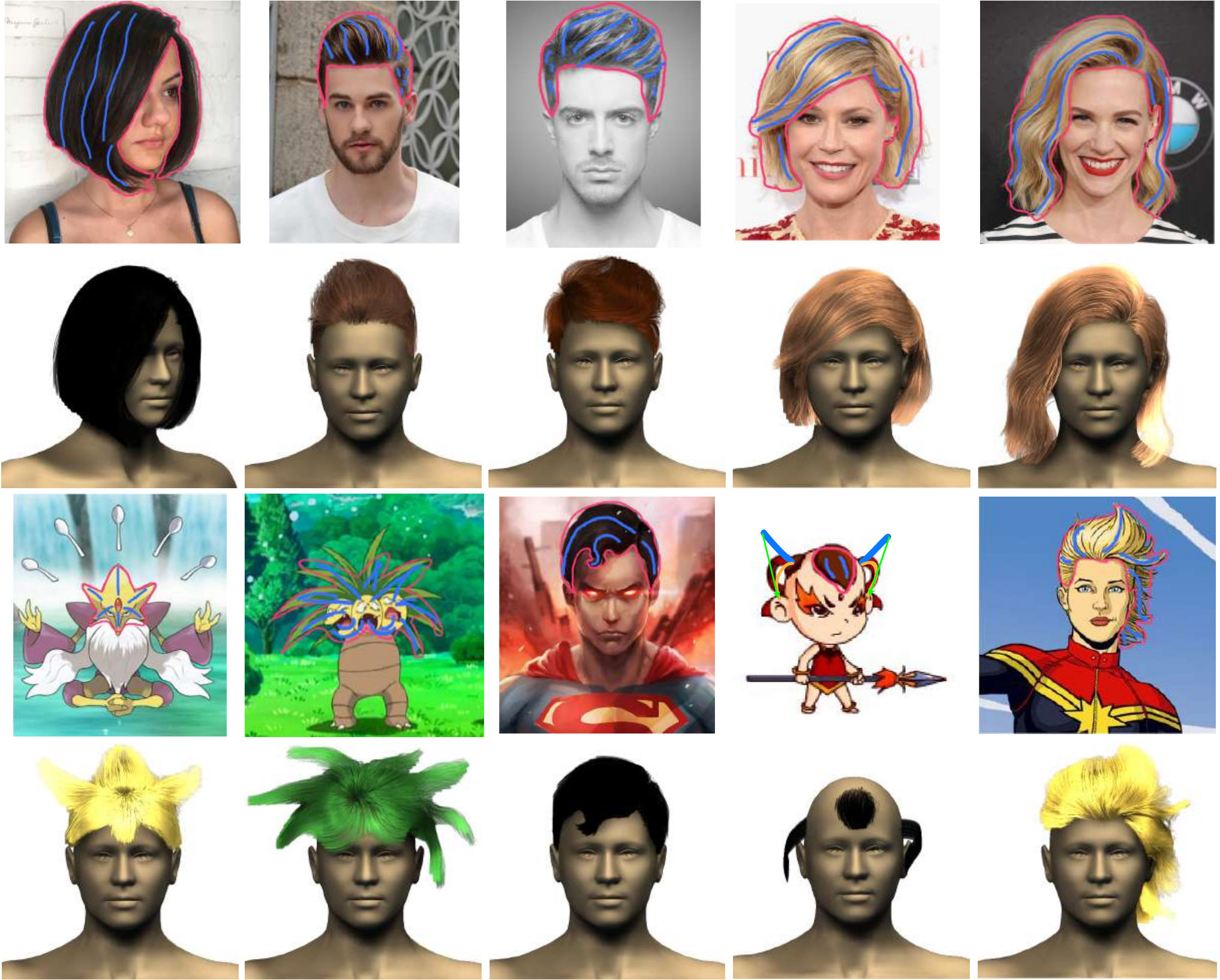}
	\caption{{Representative 3D hairstyles created by 5 novice participants (reference images and the corresponding sketches are shown atop).}}
	\label{fig:user_results}
\end{figure}

\textbf{Pilot Study.} We invited five novice users {with no 3D modeling experience} and no training with our system. We first let them get familiar with our user interface and the functions of our system. Then we supplied them with some reference Internet images and taught them how to model {a hairstyle from one of images} using our system. The training step took about 5 minutes for each participant. In the evaluation test, the users were asked to do hair modeling tasks with {a set of randomly selected images (3 for each) and try to do some artistic design. {The sketches were created by using} a Wacom touch screen with digital pen interface. We show some representative generated results and the corresponding reference images {with the input sketches overlaid} in Fig.~\ref{fig:user_results}.} The average timing statistics are reported in Table~\ref{tab:time_consuming}. The users could get a generated 3D hair model in a few seconds when they finished their drawing and clicked the ``build'' button. {Feedbacks from the participants were mainly positive but with the main concern that the response time of a few seconds is a bit long. Some of them expressed that the sketching interface in 2D is somehow limited when they wanted to deform and manipulate strands in 3D. Yet} they all expressed our \sketchhair~system can produce user-wanted hair models by simple interaction.

\begin{table}
    \renewcommand{\arraystretch}{1.3}
    \centering
    \begin{tabular}{|c|c|c|c|c|c|}
         \hline
         \bfseries Step & User Drawing & \sonet & \ovnet & \vvnet & Hair Synthesis\\
         \hline
         \bfseries Time & $\approx$ 30s & $\approx$ 0.2s & $\approx$ 0.3s & 0.2s & $\approx$ 5s\\
         \hline
    \end{tabular}
    \caption{Average performance of our various algorithmic components in the pilot study.}
    \label{tab:time_consuming}
\end{table}

\subsection{Comparisons}\label{subsec:experiment_comparison}

\begin{figure}
	\centering
	\includegraphics[width=\linewidth]{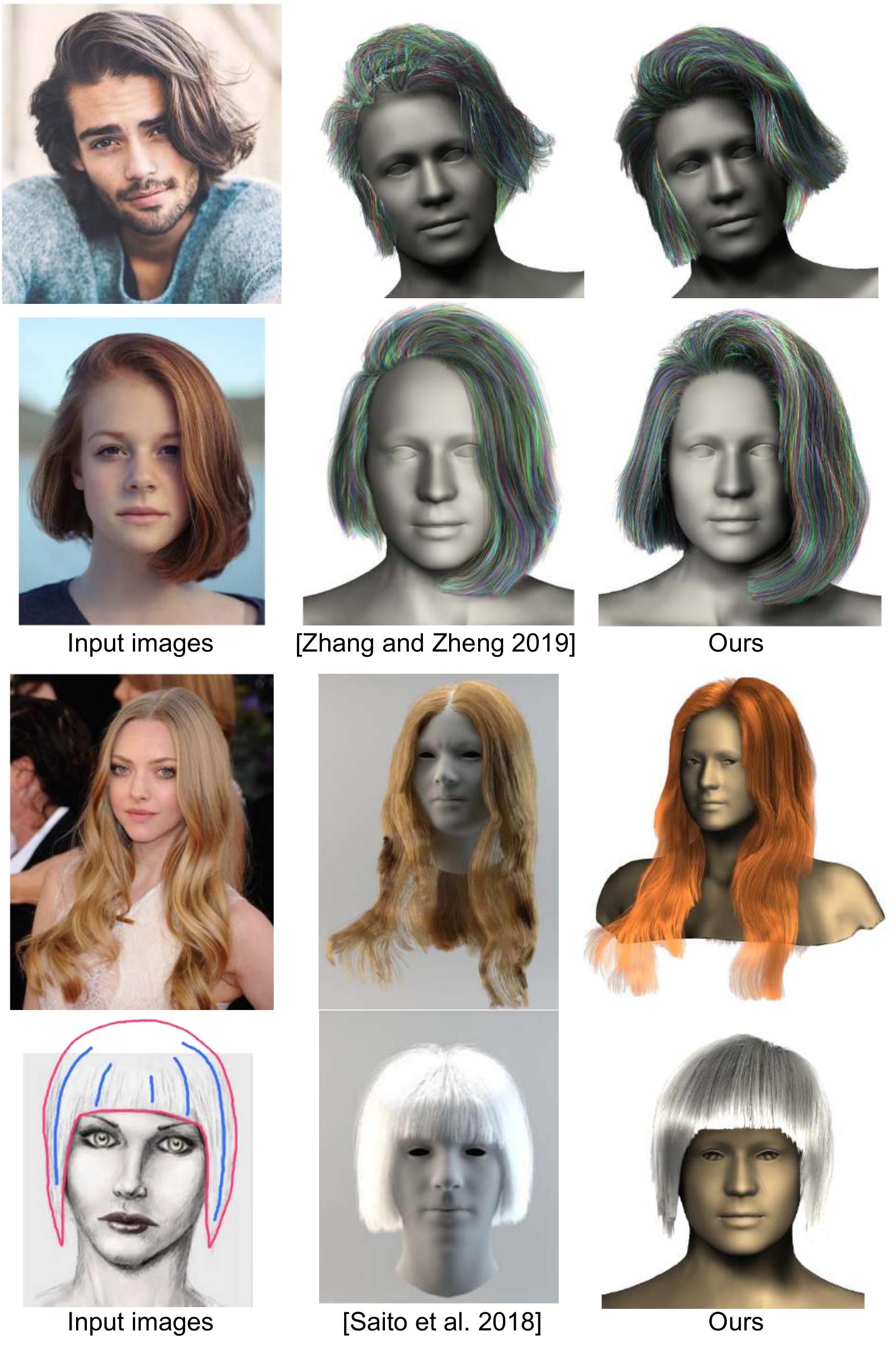}
	\caption{Comparisons of our method with {single-view} image-based hair modeling methods.}
	\label{fig:comparison_image}
\end{figure}

\begin{figure}[t!]
	\centering
	\includegraphics[width=\linewidth]{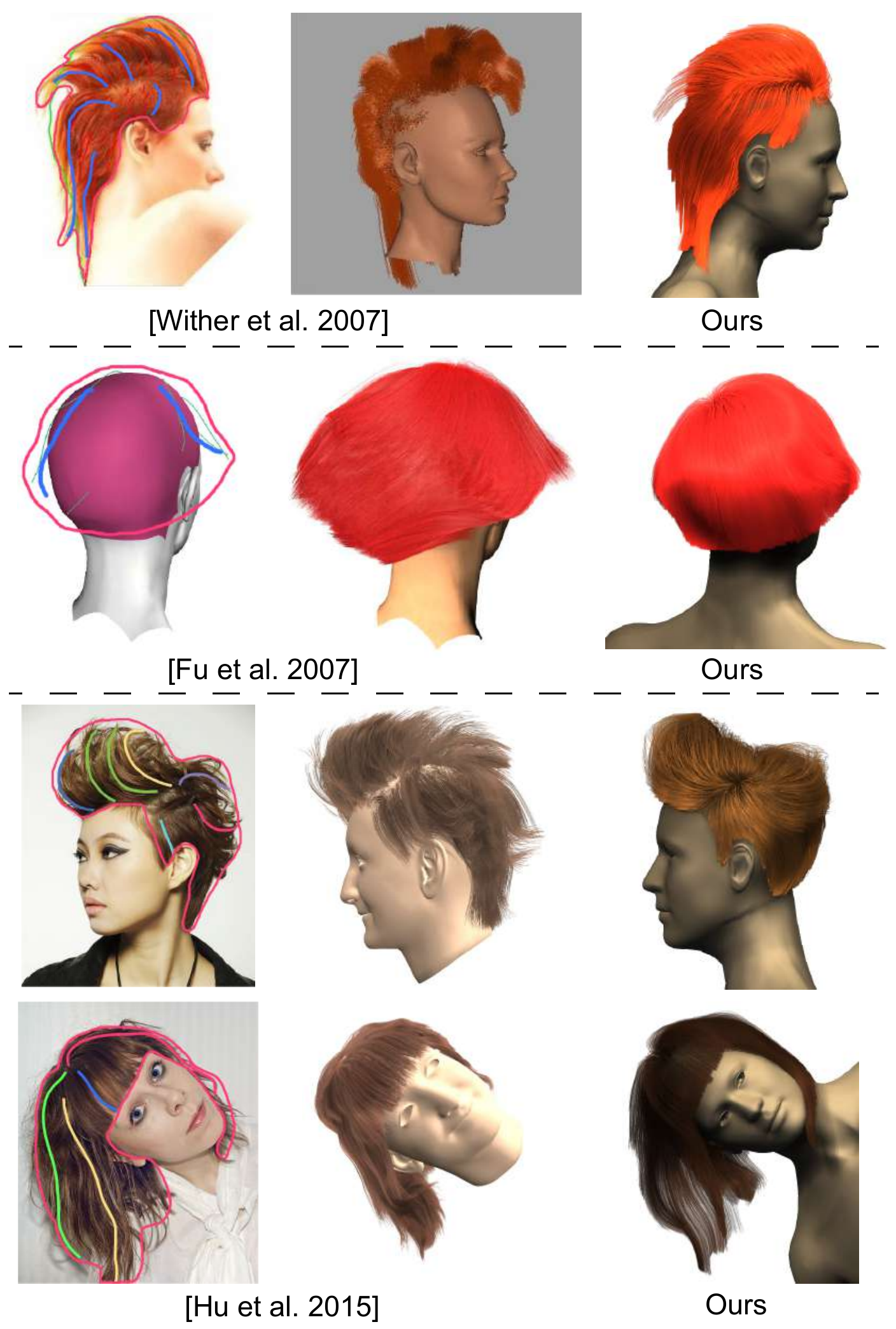}
	\caption{Comparisons of our method with previous sketch-based hair modeling techniques. In the last two rows, we use the same set of strokes as in \cite{hu2015single}.}
	\label{fig:comparison_sketch}
\end{figure}

To the best of our knowledge, our method is the first method that can model high quality 3D hair models from sets of multi-view sketches. We compare our method with the state-of-the-art single-view image-based hair modeling methods~\cite{zhang2018hair, saito20183d} and some interactive sketch-based methods~\cite{wither2007realistic, fu2007sketching, hu2015single}.

We compare our \ovnet~with \emph{HairGAN} proposed by Zhang and Zheng ~\cite{zhang2018hair}. {For fair comparison, }{we generate the hair mask and {dense} 2D orientation map from the input image as in~\cite{zhang2018hair} and feed them into both HairGAN and our \ovnet~(note that user sketches are not used here).} Our generated hair strands have higher quality even without being post-processed (in Zhang and Zheng \cite{zhang2018hair}, tedious post-processing is required for refinement). As shown in the first two rows of Fig.~\ref{fig:comparison_image}, our hair strands are {smoother and match the input images better.}

The method by Saito et al.~\cite{saito20183d} can produce 3D hair strands from cartoon images automatically. With simple or no interaction {(shown in Fig. \ref{fig:comparison_image})}, our \sketchhair~can also generate 3D hair strands referring to  cartoon images. Compared to~\cite{saito20183d}, our results can fit cartoon hairstyles better with {a small amount of user interaction.}

Earlier sketch-based hair modeling methods~\cite{wither2007realistic} rely on the physically-based hair wisp model, which cannot maintain the balance between reality and user input sketches. As the first row of Fig.~\ref{fig:comparison_sketch} shows, the result of~\cite{wither2007realistic} sometimes does not match the input strokes well, while our result is highly realistic and respects the input sketch well. We also compare our method with~\cite{fu2007sketching}, which requires user-specified 3D curves to constrain the 3D orientation field by solving a Laplacian system. In the second row of Fig.~\ref{fig:comparison_sketch}, the readers can find that our result looks more reasonable, while the hair strands in their result are twisted in some degree. What's more, our method, which uses neural network forward propagation, is much faster than solving a large sparse linear system~\cite{fu2007sketching}.

The recent data-driven sketch-based hair modeling approach by Hu et al.~\cite{hu2015single} relies on the reference images and dataset-based retrieval. This method combines different retrieved hairstyles to generate the most similar 3D hair model to the reference images. Thus, their generated hairstyles appear mixed and they cannot handle well when a target hairstyle matches none of those in the database like the examples shown in the last two rows of Fig.~\ref{fig:comparison_sketch}. Our learning-based method does not have these issues. In addition, different from~\cite{hu2015single}, our method allows users to modify hairstyles from different views, so finally, we can generate 3D hair models with very complete structures. For fair comparisons, our results shown in Fig. 14 do not use additional sketches in new views.

\subsection{Ablation Studies}\label{subsec:experiment_ablation}
In this subsection, we introduce the ablation studies to show the impact of some important algorithmic components in our system.

\textbf{\sonet. }
To test the impact of our \sonet, we have explored three alternative structures:

(i) Take the same input of \sonet, we simply diffuse the stroke direction to fill the entire hair mask region by solving a Laplace equation.

(ii) We use randomly selected 2D hair strands as the input of training without 2D sketch map pre-processing mentioned in \S\ref{sec:single_view_data_generation}. This experiment shares the same network structure with \sonet~and is trained for 100 epochs, too.

The average test errors are reported in Table~\ref{tab:mse}, where {all the methods are} compared against the same ground truth. We find that our full \sonet~achieves the best performance.

(iii) {We train our \ovnet~with 2D sketches, 2D mask images, and bust depth images directly to examine the importance of the intermediate dense 2D orientation map (i.e., not using \sonet~at all). We call this net \emph{S2VNet}.} The evaluation results are shown in Fig.~\ref{fig:ablation_single} (Left). It can be seen that without dense 2D feature maps, the details of generated 3D orientation fields are far away from the ground truth, which is consistent with our inference in \S\ref{sec:single_modeling}.

\begin{table}[b!]
    \renewcommand{\arraystretch}{1.3}
    \centering
    \begin{tabular}{|c|c|}
         \hline
         \bfseries Methods & \bfseries Mean Square Error (MSE)\\
         \hline
         \bfseries Laplacian Diffuse & 6.98\\
         \hline
         \bfseries Random {Selection} & 6.60\\
         \hline
         \bfseries \sonet & \textbf{4.01}\\
         \hline
    \end{tabular}
    \caption{Ablations study on \sonet.}
    \label{tab:mse}
\end{table}

\begin{figure}
	\centering
	\includegraphics[width=\linewidth]{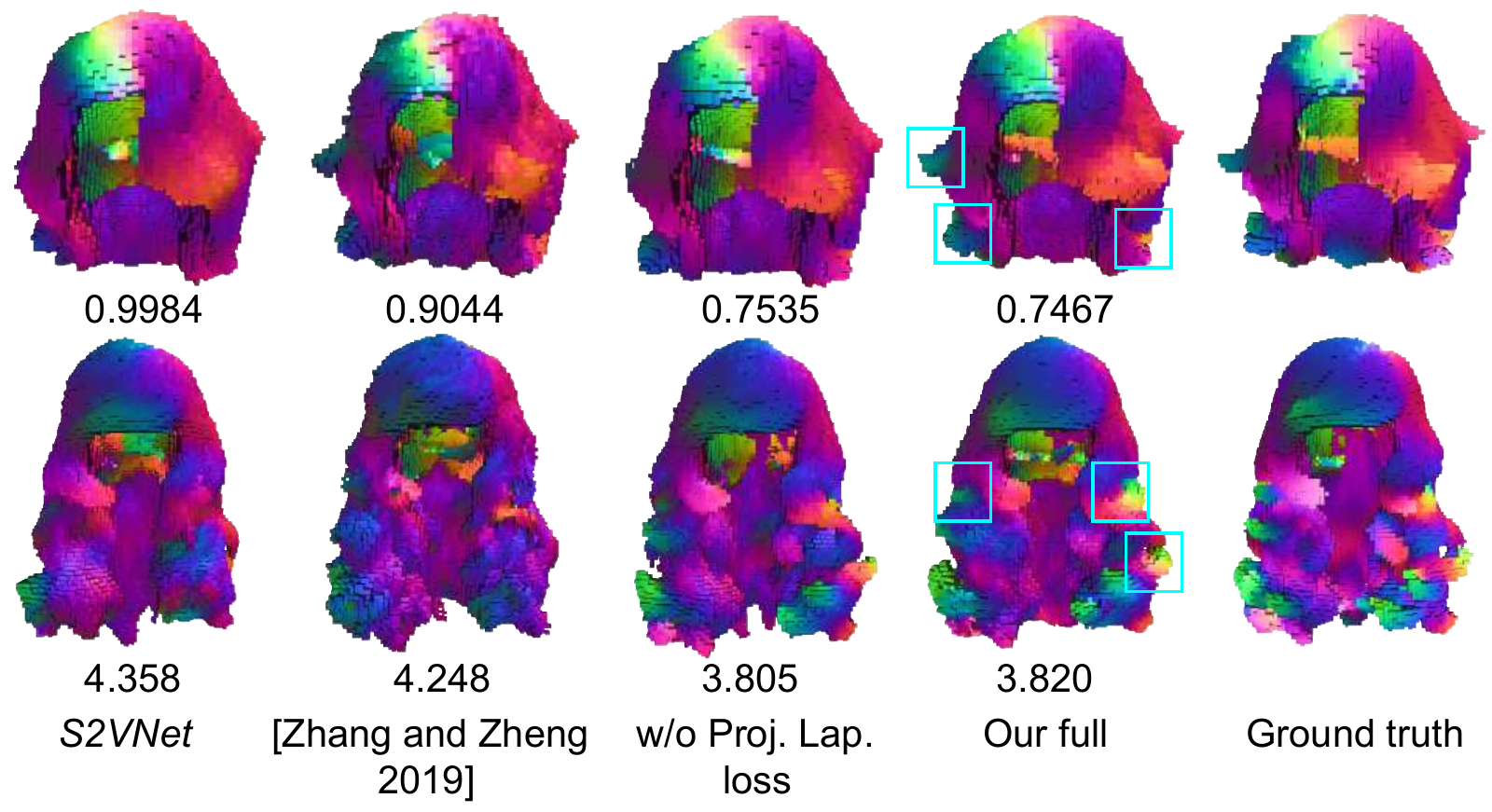}
	\caption{Ablations study on \ovnet. The projection and Laplacian losses help our results generalize well to the inputs. {The numbers indicate the MSE errors}.}
	\label{fig:ablation_single}
\end{figure}

\textbf{\ovnet~Structure and Loss Function. }
To present our improvement on 3D orientation field generation, we firstly compared our network structure with the \textit{HairGAN} proposed by Zhang and Zheng~\cite{zhang2018hair}. In this experiment, we train these two networks with the same dataset until they converged. {The $2^{nd}$ and $4^{th}$ columns of} Fig.~\ref{fig:ablation_single} show the differences between two network outputs. {The inputs to the two networks are both derived from a 3D hair model in the test set.} The readers can find that not only the errors of our network outputs are lower, {but also our synthesized 3D orientation fields are more similar to the ground truth with higher smoothness. Further, our network converges faster than that of~\cite{zhang2018hair} due to our much smaller input size.}

We also examained the effects of the projection and Laplacian losses in training. As shown in Fig.~\ref{fig:ablation_single}, {although quantitatively the errors without the projection and Laplacian loss (middle column) are {comparable to} those with our full model}, one can see visually that the results with the projection and Laplacian losses have higher curling degree, better matching with the ground truth.

\begin{figure}[t!]
	\centering
	\includegraphics[width=\linewidth]{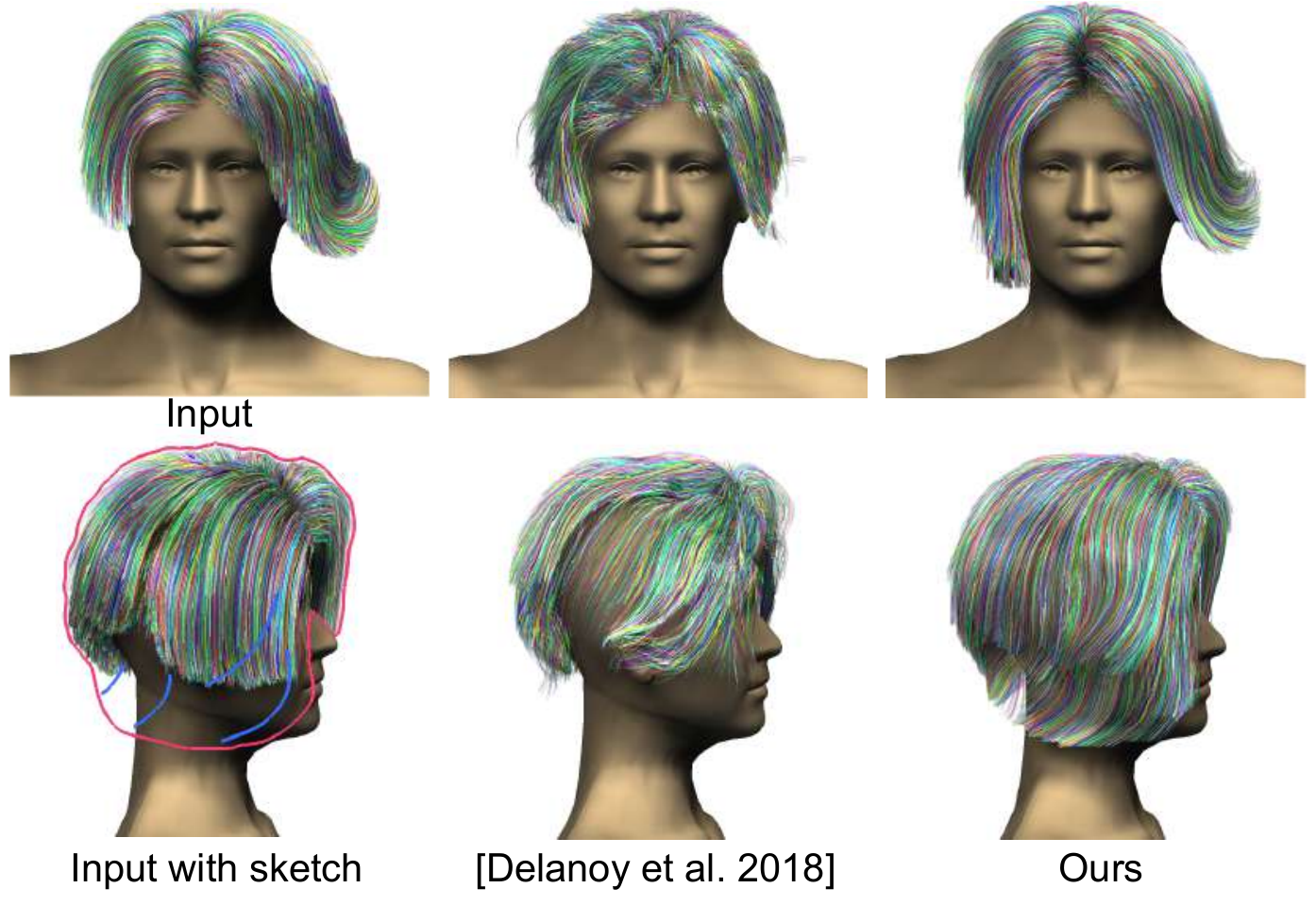}
	\caption{Ablations study on \vvnet. The method by Delanoy et al. \cite{delanoy20183d} tends to generate vector fields {whose values are inconsistent with the ones in the previous view and thus are often chaotic (the middle column). {For each method, we show the corresponding results in the previous and current views.}
	}}
	\label{fig:ablation_multi}
\end{figure}

\textbf{Volume-to-volume Generation. }
Although Delanoy et al.~\cite{delanoy20183d} present an update CNN dealing with multi-view sketch inputs to modify 3D surface models, we find this approach does not work well in our multi-view hair modeling task. We did an experiment to compare our volume-to-volume network structure with their method, which concatenates 3D volume with 2D feature maps and feeds them to a network using 2D convolutional layers only. In this experiment, we also train these two networks with the same dataset and the same loss function (we do not add the projection and laplacian loss). The readers can notice from Fig.~\ref{fig:ablation_multi} that our network performs much better than the update CNN in~\cite{delanoy20183d}: our method well preserves the features in the previous view and fills the user-specified empty mask region well.

\begin{figure}[b!]
	\centering
	\includegraphics[width=\linewidth]{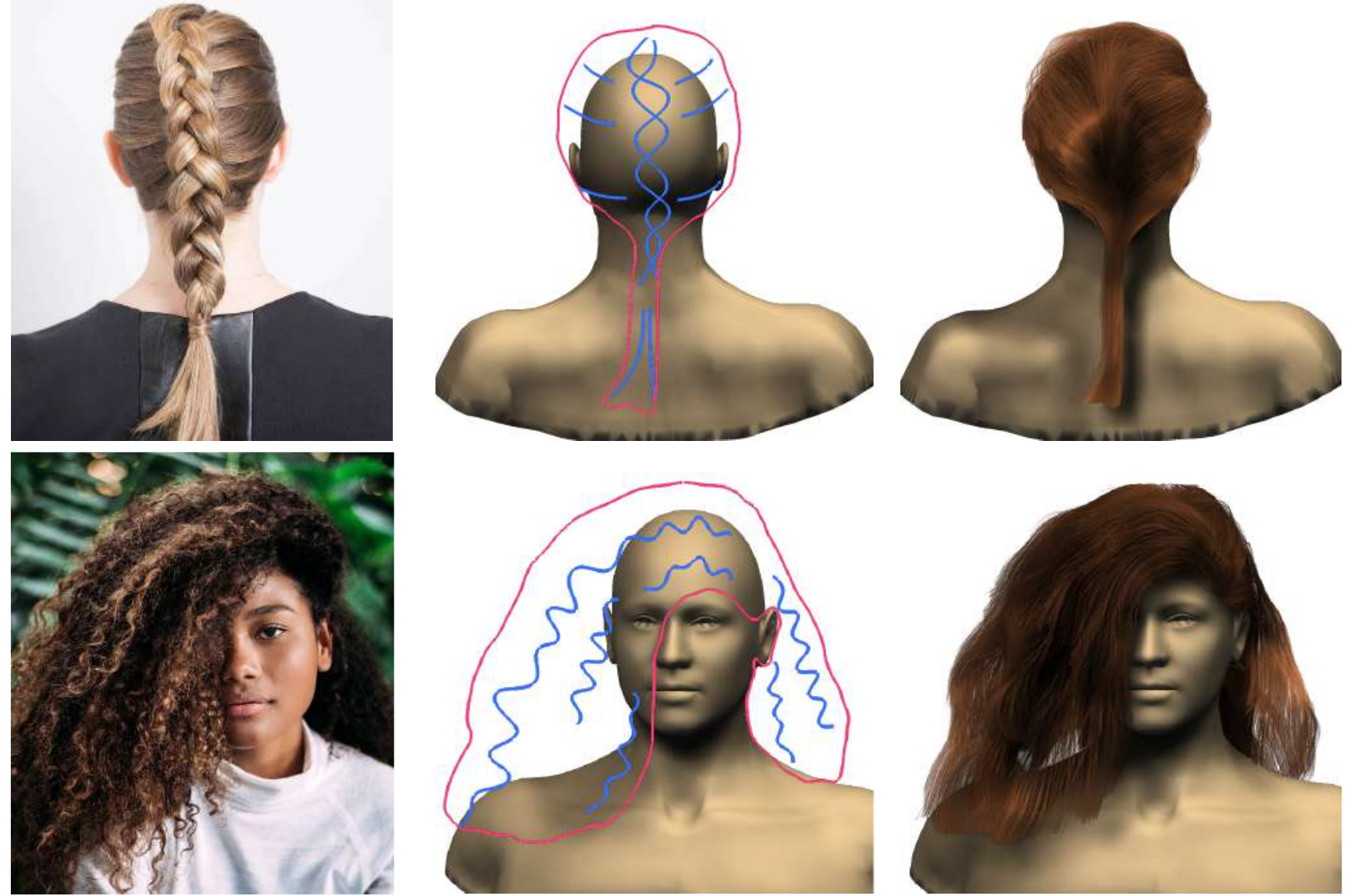}
	\caption{{Our method might fail to synthesize hair details due to either insufficient resolution of 2D or 3D orientation fields, or insufficient training data.}}
	\label{fig:failure}
\end{figure}

\subsection{Limitations}\label{subsec:experiment_limitations}
Our method has several limitations. There still exist complicated hairstyles that our \sketchhair~system cannot deal with (Fig. \ref{fig:failure}). For example, some hairstyles like Afro ({Fig. \ref{fig:failure} (Bottom)}) might have fine hair details which are difficult to handle with our pre-set resolution of 3D orientation fields. {Similar lack-of-fine-detail effects can also be observed from Fig. \ref{fig:qualitativ_results} (c) and (d). This is {a common} drawback of the current deep learning based hair modeling approaches that rely on a vector field representation \cite{saito20183d}, \cite{zhang2018hair}.} Achieving higher resolutions will bring in tremendous computational cost and need GPUs with bigger memory for training. In addition, as a learning-based approach, the effectiveness of our method is also determined by the available training data. Since we have no hairstyles like polystrip in our dataset, it is difficult for our network to learn such special shapes ({Fig. \ref{fig:failure} (Top)}).

\section{Conclusions}\label{sec:conclusion}
In this work, we have presented a novel sketching system \sketchhair~for interactive hair modeling. We showed that a deep learning based method established on a 3D hairstyle dataset is able to generate various realistic 3D hair models in accordance with 2D sketches, in single or multiple views. With a simple sketch or an image, users can get a corresponding generated 3D hair model and then {optionally}  perform further manipulations to get desired results. Our system is composed of three deep learning modules which are {carefully designed} to support seamless user interactions. The experimental results show the outstanding performance of our modeling system, and demonstrate the effectiveness and expressiveness of our proposed method. {In our current implementation, we require users to draw the hair mask in full {(even for editing in new views)} and the directed strokes to depict the hair growing direction. More intelligent interfaces or alternative interactions (for examples, those in VR) could be investigated to ease the interactions and to enable more powerful hair editing in 3D.}


%

\appendices


\ifCLASSOPTIONcompsoc
  \section*{Acknowledgments}
  We would like to thank...
\else
  \section*{Acknowledgment}
\fi


\ifCLASSOPTIONcaptionsoff
  \newpage
\fi



%

\bibliographystyle{IEEEtran}
\bibliography{Sketch_Hair}

\begin{IEEEbiography}[{\includegraphics[width=1in,height=1.25in,clip,keepaspectratio]{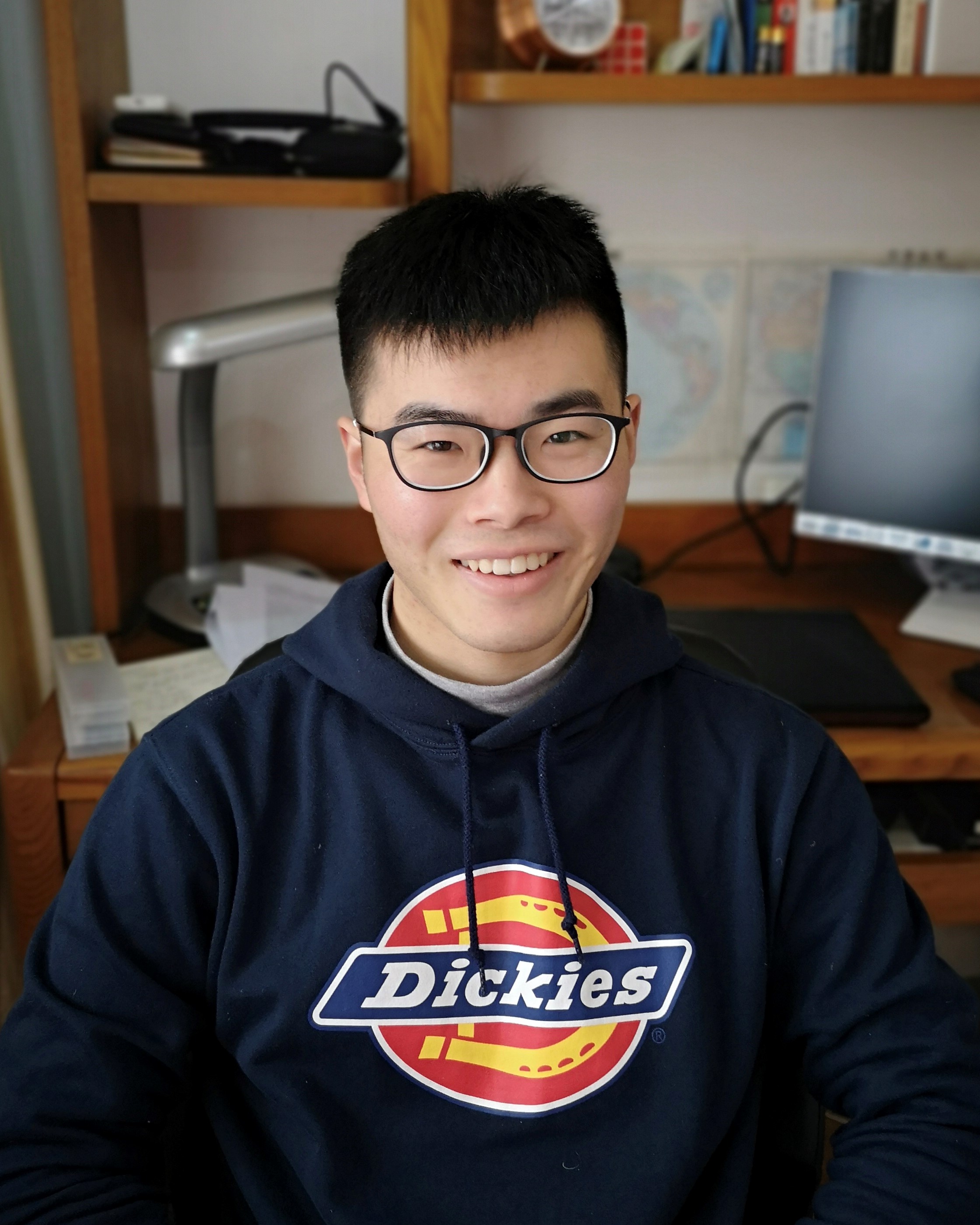}}]{Yuefan Shen} is a Ph.D. candidate at the State Key Lab of CAD\&CG, Zhejiang University. He obtained his B.S. from the School of Software Engineering at Shandong University. His research interests include image-based modeling and 3D data processing with deep learning.
\end{IEEEbiography}

\begin{IEEEbiography}[{\includegraphics[width=1in,height=1.25in,clip,keepaspectratio]{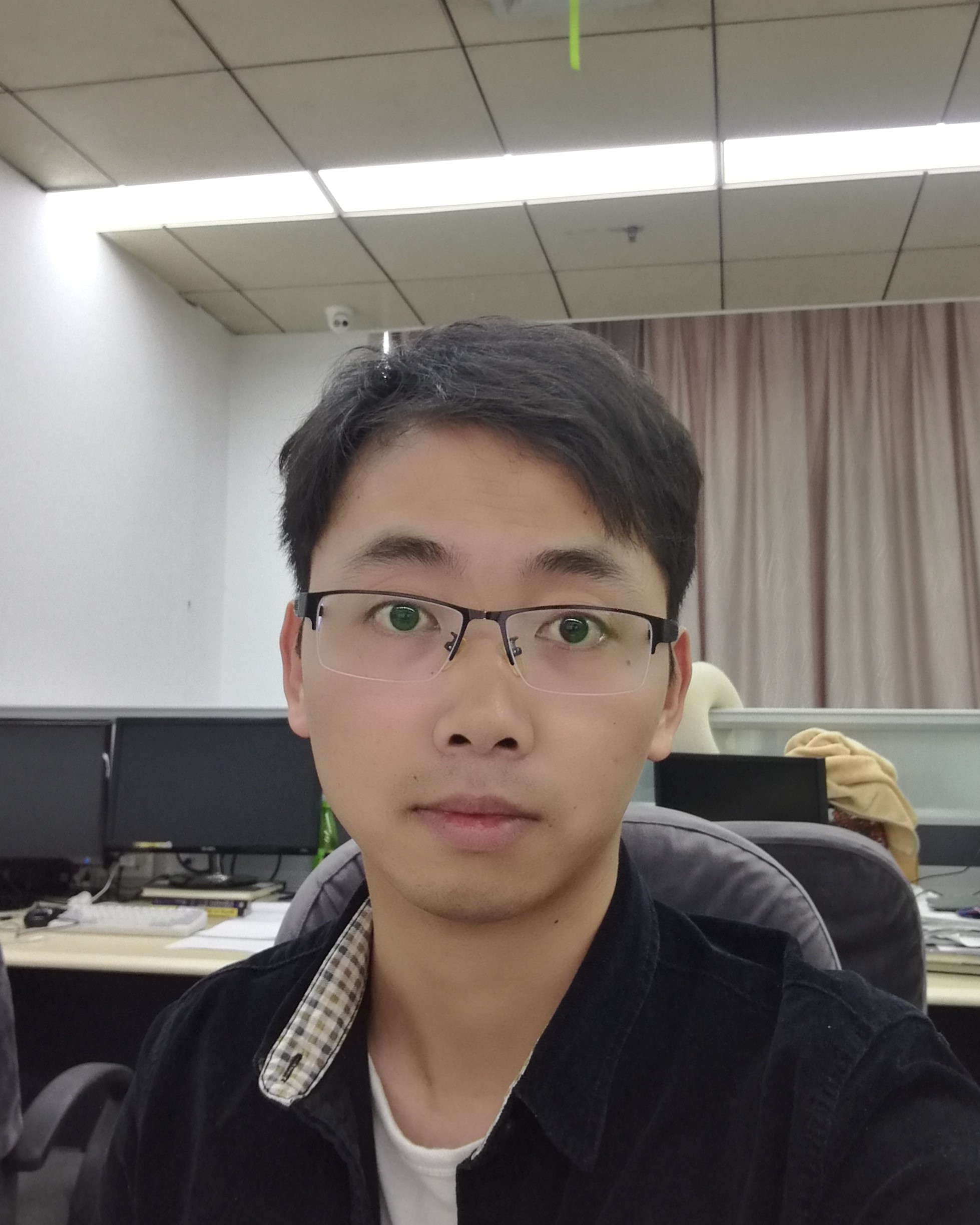}}]{Changgeng Zhang} is a Master student at the State Key Lab of CAD\&CG, Zhejiang University. He obtained his B.S. from the School of Computer Science and Technology, Huazhong University of Science and Technology. His research interests include image-based modeling and 3D data processing with deep learning.
\end{IEEEbiography}

\begin{IEEEbiography}[{\includegraphics[width=1in,height=1.25in,clip,keepaspectratio]{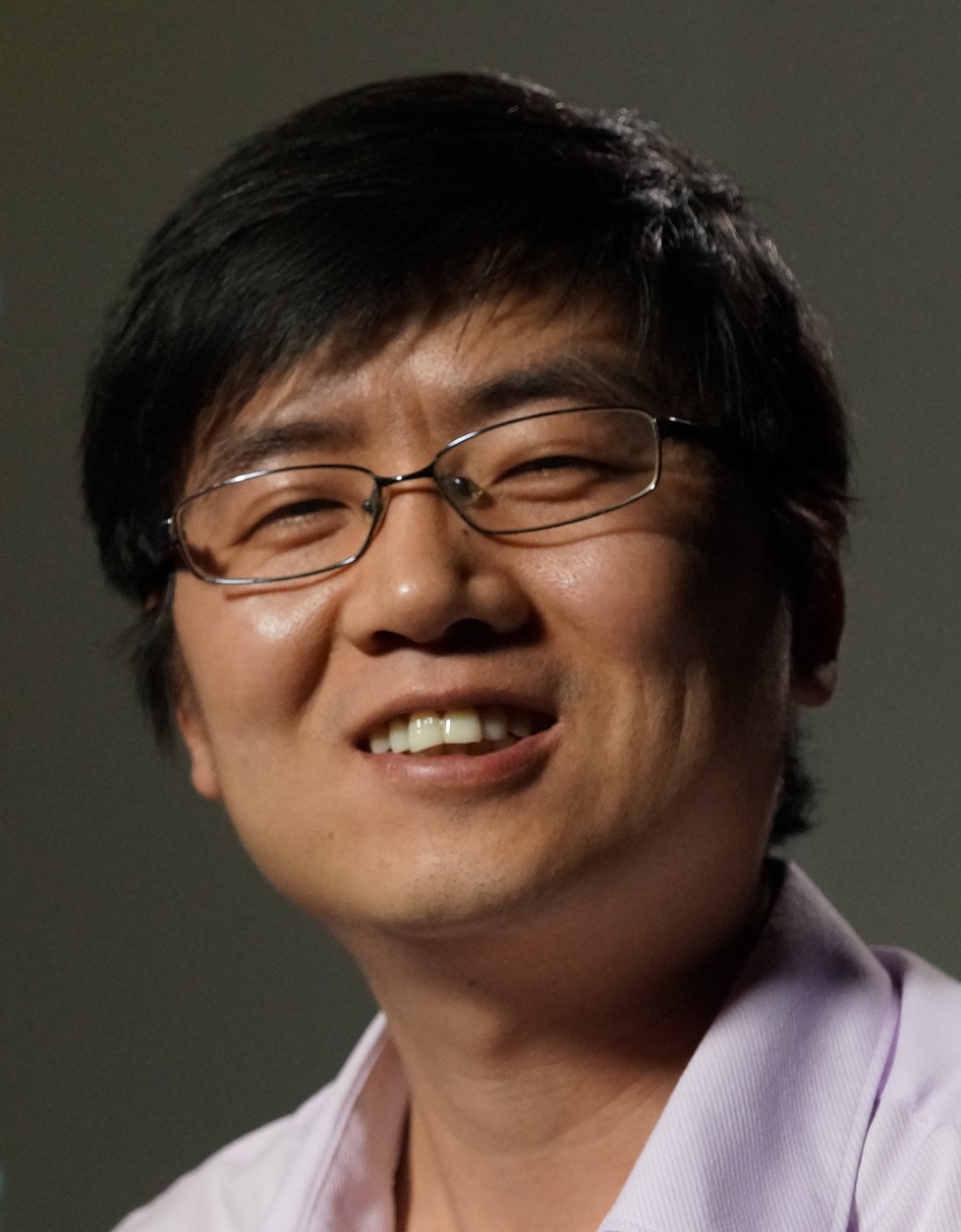}}]{Hongbo Fu} received a BS degree in information sciences from Peking University, China, in 2002 and a PhD degree in computer science from the Hong Kong University of Science and Technology in 2007. He is a Full Professor at the School of Creative Media, City University of Hong Kong. His primary research interests fall in the fields of computer graphics and human computer interaction. He has served as an Associate Editor of The Visual Computer, Computers \& Graphics, and Computer Graphics Forum.
\end{IEEEbiography}

\begin{IEEEbiography}[{\includegraphics[width=1in,height=1.25in,clip,keepaspectratio]{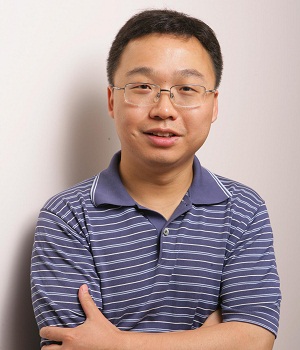}}]{Kun Zhou} is a Cheung Kong Professor in the Computer Science Department of Zhejiang University. He received his B.S. degree and Ph.D. degree in computer science from Zhejiang University in 1997 and 2002, respectively. His research interests are in visual computing, parallel computing, human computer interaction, and virtual reality. He currently serves on the editorial advisory boards of ACM Transactions on Graphics and IEEE Spectrum. He is a Fellow of IEEE.
\end{IEEEbiography}

\begin{IEEEbiography}[{\includegraphics[width=1in,height=1.25in,clip,keepaspectratio]{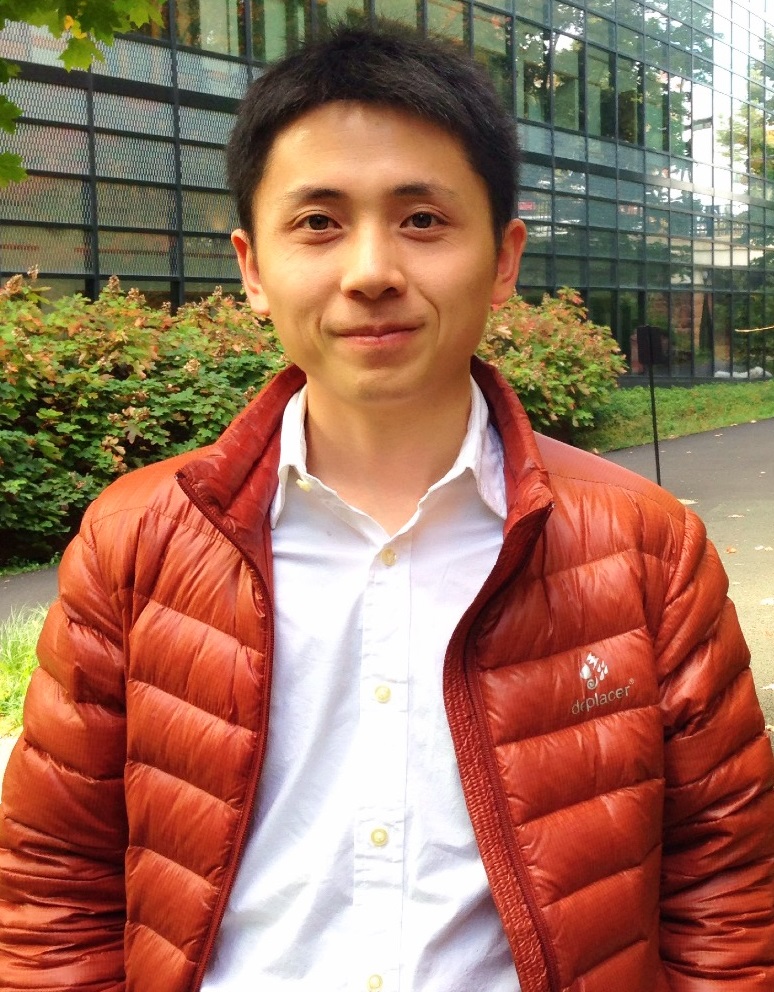}}]{Youyi Zheng} is a Researcher at the State Key Lab of CAD\&CG, Zhejiang University. He received a BS degree and an MS degree in Mathematics, both from Zhejiang University, China, in 2005 and 2007, and a PhD in Computer Science from the Hong Kong University of Science \& Technology in 2011. His research interests include geometric modeling, imaging, and human-computer interaction. He has served as an Associate Editor of The Visual Computer and Frontiers of Computer Science.
\end{IEEEbiography}

\end{document}

%% file: comments_tool/macros.tex
\usepackage{wrapfig}
\usepackage{hyphenat}
\usepackage{soul,color}
\usepackage{amsopn}
\usepackage{amsmath}
\usepackage{amssymb}
\usepackage{tabularx}
\usepackage{subfigure}

\definecolor{Red}{cmyk}{0,1,1,0}
\definecolor{Green}{cmyk}{1,0,1,0}
\definecolor{Cyan}{cmyk}{1,0,0,0}
\definecolor{Purple}{cmyk}{0.45,0.86,0,0}
\definecolor{Rosolic}{cmyk}{0.00,1.00,0.50,0}
\definecolor{Blue}{cmyk}{1.00,1.00,0.00,0}
\definecolor{Black}{cmyk}{1,0,0,1}